\documentclass{article}
\usepackage{PRIMEarxiv}
\usepackage{amsmath, amssymb}
\usepackage{graphicx}
\usepackage{xcolor}
\usepackage{graphicx}
\usepackage{amsmath,amsfonts,amssymb,mathrsfs}
\usepackage{caption}
\usepackage{booktabs}
\usepackage{subcaption}
\usepackage{hyperref}
\usepackage[utf8]{inputenc}
\usepackage{algorithm}
\usepackage[noend]{algpseudocode}
\usepackage{tabularx} % à ajouter dans le préambule
\title{InfinityEBSD : Metrics-Guided Infinite-Size EBSD Map Generation With Diffusion Models}
\author{
  Sterley Labady, Youssef Mesri, Daniel Pino Munoz, Baptiste Flipon, Marc Bernacki \\
  Mines Paris, PSL University\\
  Centre for material forming (CEMEF), UMR CNRS\\
  06904 Sophia Antipolis, France\\
  \texttt{marc.bernacki@minesparis.psl.eu, sterley.labady@minesparis.psl.eu} \\
}

\begin{document}
\maketitle

\begin{abstract}
Materials performance is deeply linked to their microstructures, which govern key properties such as strength, durability, and fatigue resistance. EBSD is a major technique for characterizing these microstructures, but acquiring large and statistically representative EBSD maps remains slow, costly, and often limited to small regions. In this work, we introduce InfinityEBSD, a diffusion-based method for generating monophase realistic EBSD maps of arbitrary size, conditioned on physically meaningful microstructural metrics. This approach supports two primary use cases: extending small experimental EBSD maps to arbitrary sizes, and generating entirely new maps directly from statistical descriptors, without any input map. Conditioning is achieved through eight microstructural descriptors, including grain size, grain perimeter, grain inertia ratio, coordination number and disorientation angle distribution, allowing the model to generate maps that are both visually realistic and physically interpretable. A patch-wise geometric extension strategy ensures spatial continuity across grains, enabling the model to produce large-scale EBSD maps while maintaining coherent grain boundaries and orientation transitions. The generated maps can also be exported as valid Channel Text Files (CTF) for immediate post-processing and analysis in software such as MTEX or simulation environments like DIGIMU\textsuperscript{\textregistered}. We quantitatively validate our results by comparing distributions of the guiding metrics before and after generation, showing that the model respects the statistical targets while introducing morphological diversity. InfinityEBSD demonstrates that diffusion models, guided by physical metrics, can bridge the gap between  synthetic and realistic materials representation, paving the way for future developments such as 3D realistic microstructure generation from 2D data.
\end{abstract}


\section{Introduction}
\label{introduction}

Materials play a critical role in modern industries, particularly in sectors such as aerospace, automotive, and renewable energy. The design and optimization of materials are intrinsically linked to their microstructures, which directly influence their performance and durability. Grain orientation and size, boundary connectivity (number of neighbors, disorientation) shape the macroscopic behavior of materials. Despite their importance, the acquisition of high-quality microstructures data, remains a significant challenge. Even with the increasing availability of advanced imaging techniques, including Electron Backscatter Diffraction (EBSD) \cite{Schwartz2009} and Focused Ion Beam - Scanning Electron Microscopy (FIB-SEM) \cite{Giannuzzi2005}, most microstructure datasets remain expensive to obtain, and constrained in size. These constraints are particularly visible in 2D EBSD maps acquisitions \cite{Randle2009}, where they are often limited to small regions, capturing just a few dozen to a few hundred grains. While these are sufficient for qualitative inspection, they are rarely enough to provide statistical and physical representativeness. This limitation can lead to biased analysis, especially in materials with heterogeneous microstructures. Moreover acquiring larger EBSD maps, which offer a better statistical representation, is slow and expensive, both in terms of microscope time and post-processing effort \cite{Doddapaneni2025}. These constraints have increased interest in generating EBSD maps of arbitrary size through data-driven methods that maintain realistic microstructural characteristics.\\

Given the practical constraints of experimental microstructure acquisition, computational generation has emerged as a valuable and complementary strategy. By simulating microstructures \cite{Srolovitz1984, Tourret2022, Moelans2008}, instead of capturing them physically, researchers gain access to a flexible and scalable approach that can significantly reduce cost and experimentation time. Synthetic microstructures make it possible to explore conditions that would be too time-consuming or difficult to reproduce experimentally. In addition, they provide a controlled baseline for assessing and refining experimental techniques.

Synthetic microstructure reconstruction is not new. Traditional approaches to microstructure reconstruction often rely on statistical and physical models. Techniques such as correlation functions \cite{Torquato2002} and geometric approximations using tessellation methods (e.g., Voronoi \cite{Okabe2000} and Saltykov \cite{Saltykov1967}) have been widely used to generate polycrystalline microstructures \cite{Hitti2012}. However, they tend to produce oversimplified grain morphologies with simplistic shapes and are rarely able to capture real-world microstructure details and boundary networks. More recently, machine learning models such as Convolutional neural networks (CNN) \cite{Lecun1998}, Generative Adversarial Networks (GAN) \cite{Goodfellow2014}, and Denoising Diffusion Probabilistic Models (DDPM) \cite{Ho2020, SohlDickstein2015} have been applied with success to microstructure generation \cite{Lee2024, Kench2021, Bostanabad2020, Hoffman2025, Dureth2023, Phan2024}. However, most of these models treat microstructures as grayscale images \cite{Dureth2023, Lee2024, Kench2021, Phan2024}. As a result, they may produce plausible-looking textures that lack physical meaning.\\

In this work, we adopt the view that generative models should not create random microstructures. Instead, these models need to be guided or constrained by physically meaningful information such as grain size, shape, and orientation distribution. To date, there is no established method that can generate EBSD maps of arbitrary size while ensuring consistency with a complete set of microstructural metrics. These include grain size, grain perimeter, grain inertia ratio, coordination number, grain boundary disorientation and the three Euler angles that describe crystal orientation. The ability to generate EBSD maps that match given metrics is crucial for producing synthetic data that is not only visually realistic but also physically meaningful for simulations or analysis.

\begin{figure}[ht]
    \centering
    \includegraphics[width=0.9\linewidth]{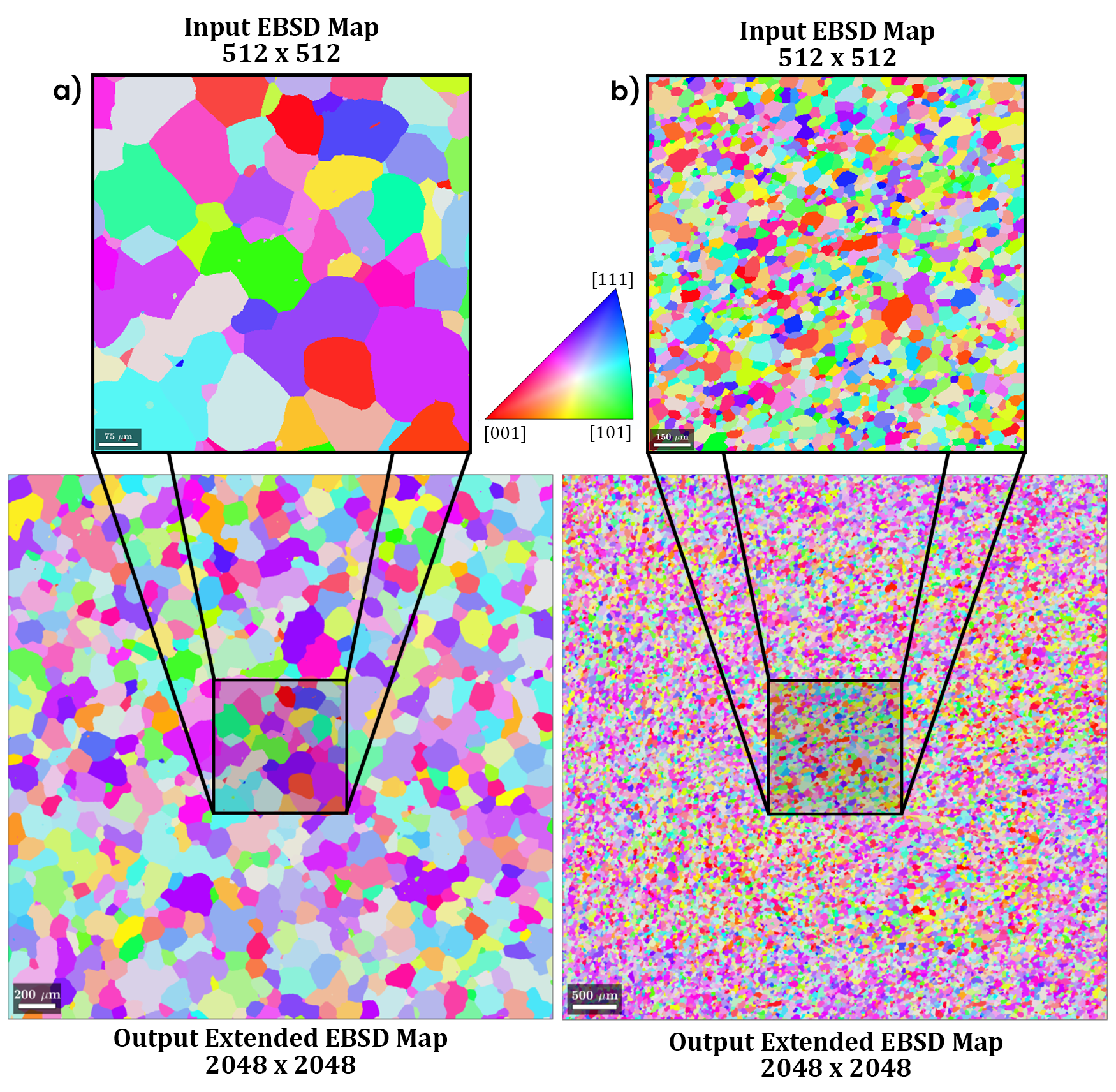}
    \caption{
        Overview of the proposed method for extending EBSD maps from small ones (512×512 cells) to large generated maps (2048×2048 cells). The generated microstructures follow statistical and physical characteristics of the input microstructures. Map (a) corresponds to Inconel 718 (In718), a nickel-based superalloy and Map (b) corresponds to an austenitic stainless steel (316L).
    }
    \label{fig:graphical_abstract}
\end{figure}

In this work, we present InfinityEBSD, as shown in Figure \ref{fig:graphical_abstract}, a new method to generate large EBSD maps of arbitrary size, conditioned on microstructural metrics, using DDPM trained in the latent space of a Variational AutoEncoder (VAE) \cite{Kingma2014}. We extend small EBSD maps of arbitrary size input, into large ones of arbitrary size output, preserving both the morphological realism and the statistical properties of the input map. Additionally, the model can generate entirely new EBSD maps from scratch by conditioning on a set of target microstructural metrics, without requiring any input map.\\

When dealing with high-resolution data such as EBSD maps, a common problem with DDPM is their computational consumption \cite{Rombach2022, Jiang2025, Zhao2025, Zhao2023}. That is why we address this issue by shifting the generation process to a compressed latent space. Rather than relying on raw experimental Euler angles values, we first encode EBSD maps into a compact 4D latent space learned by a dedicated VAE \cite{Rombach2022}. This separation allows the model to focus its generative capacity on meaningful map content while discarding imperceptible noise or repetitive information. This results in substantial gains in efficiency without a noticeable loss in detail or structure. A conditional UNet \cite{Ronneberger2015} then learns to denoise these latents, conditioned on microstructural metrics encoded as continuous embeddings. As it will be illustrated, that guided DDPM actually are capable of generating high quality EBSD maps.\\

InfinityEBSD also introduces a geometric extension mechanism for generating maps of arbitrary size. Unlike standard inpainting and outpainting \cite{Pathak2016} methods that work within fixed-size contexts, our method generates a series of EBSD patches to preserve spatial continuity across grains. Our approach follows a divide-and-conquer strategy: instead of generating entire EBSD maps in a single step, we break them down into smaller patches. The model is trained to synthesize these local regions independently, and the final EBSD map is reconstructed by stitching the generated patches together. In addition, the generated maps are exportable as valid Channel Text Files (CTF) \cite{OxfordAZtecManual}, making them immediately usable by tools like MTEX \cite{Bachmann2010} and DIGIMU\textsuperscript{\textregistered} \cite{Digimu,Digimu1,Bernacki2024}.\\

To assess whether the generated EBSD maps preserve physical plausibility, we compared the statistical distributions of the key microstructural metrics between the original maps and their extended versions. This validation step ensures that the model does not generate simply random microstructures, but rather respecting the underlying statistical and physical structure of the given metrics embeddings. 

This work makes the following contributions:
\begin{itemize}
    \item A diffusion-based method for EBSD maps generation of arbitrary size conditioned on microstructural metrics, is introduced. To the best of our knowledge, this work presents the first use of DDPM for metrics-guided EBSD maps generation.
    \item A patch-wise geometric extension strategy, enabling generation of large EBSD maps with spatial continuity, is proposed. This method also allows the extension of input maps of arbitrary size.
    \item The physical relevance of the generated maps is demonstrated by exporting them as CTF, analyzing their geometric and physical statistics and making comparison of microstructural statistics between the input microstructures and the generated ones.
\end{itemize}

\section{Background}
\label{background}

\subsection{EBSD Maps}
\label{ebsd}

EBSD \cite{Schwartz2009} is a widely used SEM technique that provides crystallographic orientation information. They are valuable for characterizing microstructures in polycrystalline materials. When an electron beam hits a tilted crystalline surface, backscattered electrons form Kikuchi patterns \cite{Nishikawa1928} that can be decoded to obtain the crystal orientation at each scanned point. For monophase materials, the resulting data is commonly organized into three-dimensional orientation maps, where each cell represents the local crystal orientation, typically expressed in Euler \cite{Nolze2015} angles $(\phi_1, \Phi, \phi_2)$ relative to the sample coordinate system.\\

EBSD maps are stored using standardized text-based formats. One of the most commonly used is the CTF format, originally developed by the software platform Channel 5 by HKL Technology, later acquired by Oxford Instruments \cite{OxfordAZtecManual}. In this work, we rely on CTF files as input and output format, which ensures compatibility with established software such as MTEX for post-processing and analysis.

\subsection{Denoising Diffusion Probabilistic Models}
\label{ddpm}

DDPM are a class of generative models that have shown strong capabilities for producing high-dimensional structured data through a two-steps stochastic process: Forward Process (FP) and Reverse Process (RP). This method was originally proposed by Sohl-Dickstein et al. \cite{SohlDickstein2015}, drawing inspiration from nonequilibrium thermodynamics. The idea is to gradually transform an arbitrary data distribution into a simple, tractable distribution by successively adding noise in small increments. This is known as the FP, following a Markov chain, as defined in \cite{Ho2020}. Once this transformation has been learned, a RP is trained to invert the noise trajectory and reconstruct samples from the data distribution. As defined in \cite{Ho2020}, this generative process also follows a Markov chain.\\

In practice, a simplified training objective proposed by Ho et al. \cite{Ho2020} focuses on learning the added noise rather than directly learning the distribution parameters, leading to a stable and efficient training. Further details on the thermodynamic formulation of diffusion models and their theoretical properties can be found in the original work of Sohl-Dickstein et al. \cite{SohlDickstein2015}, while practical implementation are discussed extensively in \cite{Ho2020, Rombach2022}.\\

\begin{figure}[H]
    \centering
    \includegraphics[width=0.9\linewidth]{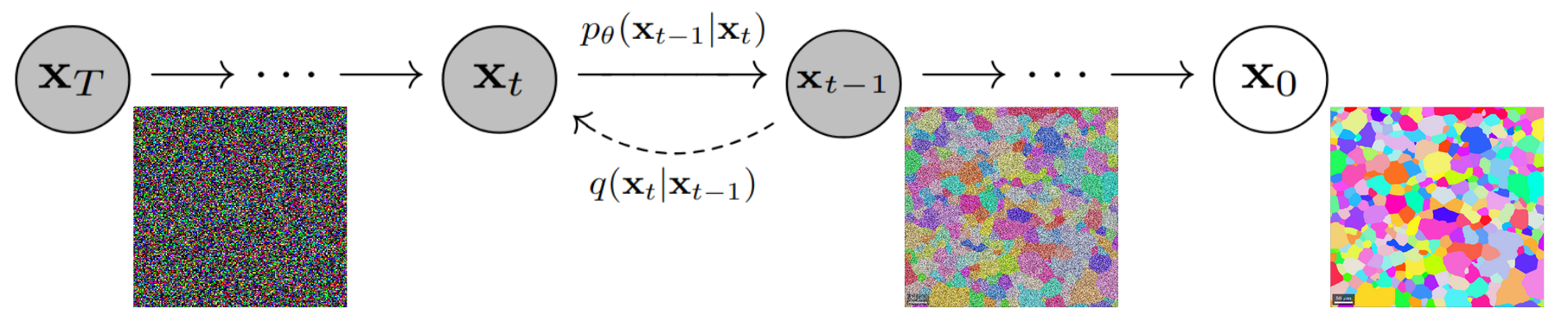}
    \caption{Overview of the denoising diffusion process, the RP. Previously the FP gradually corrupts a clean EBSD map into noise by sequentially adding Gaussian perturbations until it becomes fully noisy. Then, as the figure shows, in the RP we iteratively recover the original structure through denoising steps.}
    \label{fig:diffusion}
\end{figure}

DDPM are very powerful for microstructure generation tasks and also maps extension where it is important to preserve local continuity. Figure \ref{fig:diffusion} shows how the diffusion process works on EBSD maps: starting from a clean map, where noise is previously added in the FP, and then removed by a neural network in the RP to recover the structure.

\subsection{Attention}
\label{attention}

The attention mechanism was initially introduced in the context of natural language processing \cite{Bahdanau2016} to allow models to focus on specific parts of an input sequence when generating an output. It gained wider adoption with the development of the Transformer architecture \cite{Vaswani2017}, which replaced recurrence entirely with self-attention layers. It combines multi-head self-attention with feed-forward networks, layer normalization, and residual connections.\\

Originally designed for natural language processing, the flexible design of the Transformer architecture have made it applicable in other fields like image generation \cite{Dosovitskiy2021}. In the context of microstructure generation, attention serves a critical purpose. Instead of aligning natural language tokens, we align spatial regions of microstructural representations with metrics signals such as grain size distribution, grain orientation and disorientation distribution. These metrics are encoded into latent embeddings, which act as conditioning inputs. Through cross-attention layers, the generative model learns to align these conditioning vectors with the latent of EBSD map being synthesized. This mechanism provides the model with an internal ability to attend to specific statistical descriptors when generating or extending microstructures. As a result, the network learns to embed high-level physical constraints directly into the generation process, rather than merely hallucinating random EBSD maps regions.

\subsection{Residual Learning}
\label{residual}

Deep neural networks typically improve in performance with increasing depth, but training very deep architectures often leads to degradation where additional layers cause higher training error and become harder to optimize. Residual learning, introduced by He et al. \cite{He2016}, addressed this problem by reformulating the layers to learn residual functions instead of direct mappings. As defined in \cite{He2016}, rather than approximating a function $H(x)$, the model learns a residual function $F(x) = H(x) - x$, such that the original function becomes: $H(x) = F(x) + x$. This simple powerful idea allows gradients to flow more easily through the network during backpropagation, mitigating the vanishing gradient problem and enabling the training of much deeper architectures.\\

The core of residual learning lies in the shortcut (or skip) connection that bypasses one or more layers. It allows the model to retain information and facilitates gradient flow during training. Residual blocks became the foundation of the ResNet architectures \cite{He2016}, which demonstrated state-of-the-art performance on various computer vision tasks while significantly increasing the number of trainable layers without degradation.

\section{Method}
\label{method}

\subsection{Data Processing}
\label{dataprocessing}

\subsubsection{EBSD Maps}
\label{dataprocessingebsd}

\begin{figure}[H]
    \centering
    \includegraphics[width=0.6\linewidth]{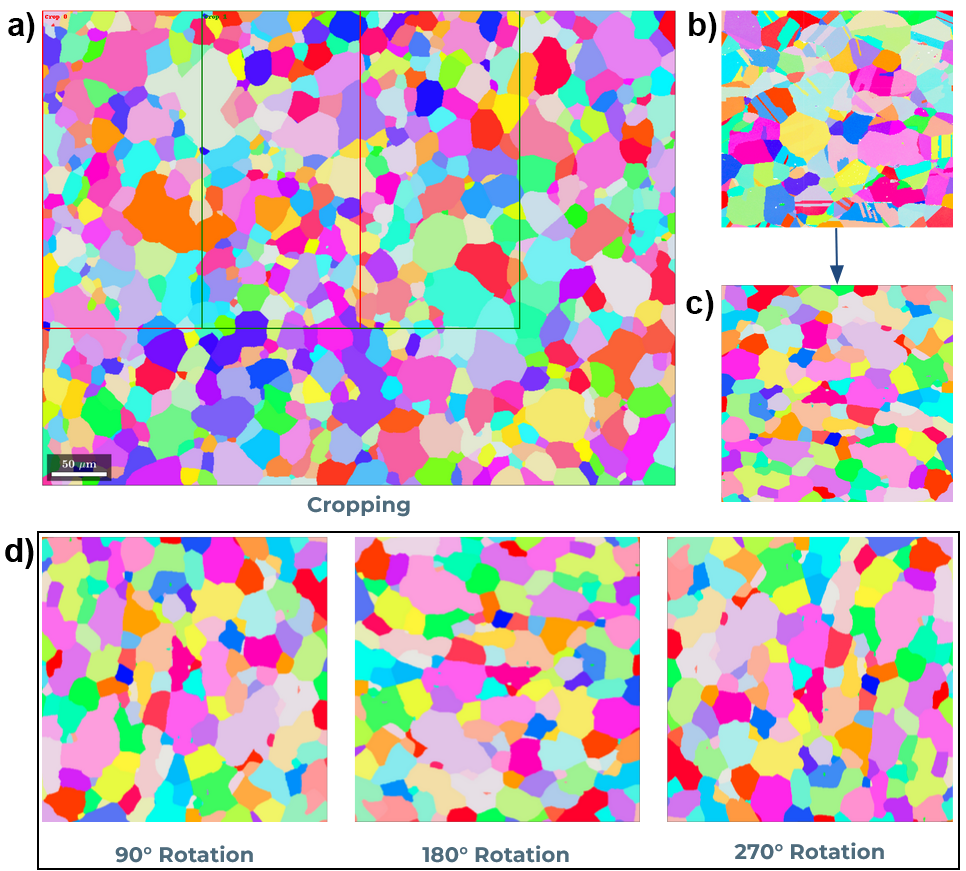}
    \caption{Illustration of the EBSD data processing pipeline. (a) Cropping of the original EBSD map into patches of $512\times512$ cells. (b) Example of a raw EBSD map. (c) The same map from (b) after pre-processing, including twin removal and the application of disorientation angle and grain size thresholds. (d) Examples of data augmentation applied to the patches, showing 90°, 180°, and 270° rotations.}
    \label{fig:datasetprocessing}
\end{figure}

All EBSD data used in this work come from fully recrystallized microstructures. Limited intragranular crystallographic orientation gradients are present but depending on the considered material annealing twins ($\Sigma3$) may be present. As shown in Figure \ref{fig:datasetprocessing}, all raw EBSD maps were pre-processed. For each EBSD map: (i) a disorientation angle threshold of 10° is used to define high-angle grain boundaries, (ii) a minimum grain area threshold is set to 10 cells to prevent small misindexed areas to be considered as grains. To decrease the complexity of raw data and due to limited intragranular orientation gradients each cell of a grain is assigned the grain mean orientation. $\Sigma3$ twins are merged and the orientation of the parent grain is assigned to the merged twin. This results in a twins free microstructure with one orientation per grain. This process eliminate acquisition noise while keeping the overall consistency of the microstructure, resulting in cleaner and more consistent maps for training.\\

To increase training diversity and allow the model to generalize, from each pre-processed EBSD map, we extract patches of size $512\times512$ cells, using a sliding window approach with 50\% overlap. As illustrated in Figure \ref{fig:datasetprocessing}, this process is followed by a series of data augmentations. Each patch is transformed via rotations (90°, 180°, 270°) and flipped horizontally and vertically. Additional combinations of flip-and-rotate are included : Horizontal Flip + 90° Rotation and Vertical Flip + 90° Rotation. These transformations expose the model to different symmetry-equivalent configurations of the same microstructure.\\ 

EBSD maps encode orientation using three Euler angles $(\phi_1, \Phi, \phi_2)$, relative to the sample coordinate system. To ensure consistent numerical scaling during neural network training, each angle channel was linearly normalized to the range $[-1, 1]$. Specifically, the domains $[0^\circ, 360^\circ]$ for $\phi_1$ and $\phi_2$, and $[0^\circ, 180^\circ]$ for $\Phi$, were rescaled accordingly.\\ 

Each EBSD map in the dataset is finally represented as a 3-channel tensor $\mathbf{X} \in \mathbb{R}^{3 \times 512 \times 512}$. The three channels correspond to the normalized Euler angles.

\subsubsection{Metrics}
\label{dataprocessingmetrics}

To guide the generative model toward producing physically realistic microstructures, we condition the diffusion process using a set of physically meaningful metrics that capture essential statistical features of each EBSD map. These metrics describe the distributions of several properties. We defined a set of 8 metrics : Grain size Distribution, Grain Perimeter Distribution, Grain Coordination Number Distribution, Grain Disorientation Distribution, Grain Inertia Ratio Distribution, and the three Grain Euler Angles $(\phi_1, \Phi, \phi_2)$ Distribution.\\ 

Grain size is estimated as the diameter of the equivalent circle associated with each grain’s area, and is expressed in cells, which refers to the unit obtained by dividing the physical dimensions by the acquisition step size, which represents the spatial resolution of the EBSD scan. The grain perimeter metric quantifies the length of each grain’s boundary, also in cells, reflecting morphological complexity. The inertia ratio describes the shape anisotropy of grains by computing the ratio between the largest and smallest eigenvalues of the grain’s inertia tensor, producing a unitless scalar $R = \lambda_{\text{max}} / \lambda_{\text{min}}$. The coordination number is determined by counting unique adjacent grains per grain, providing a discrete integer measure of local connectivity. Grain disorientation angle is calculated across boundaries taking into account the crystalline symmetry of the polycrystal. Finally, each grain’s crystallographic orientation is described by three Euler angles $(\phi_1, \Phi, \phi_2)$, measured in degrees.\\ 

For each metric, we transform the raw scalar values computed over all grains in a map into a discrete probability distribution. This is done by constructing a normalized histogram, which estimates the probability density function $P(m)$ of the metric $\mathbf{C}$ across the sample:
\begin{equation}
P(m) = \frac{1}{N} \sum_{i=1}^{N} \delta \bigl(m - m_i \bigr)
\end{equation}
where $N$ is the number of grains, $m_i$ is the value of the metric for the $i$-th grain, and $\delta(\cdot)$ is the Dirac delta function acting as a counting kernel for histogram construction. The resulting distribution is then rescaled and embedded into a fixed-size vector of length 1024. Finally, the vectors of all 8 metrics are concatenated to form a conditioning tensor $\mathbf{C} \in \mathbb{R}^{8 \times 1024}$.\\

A crucial step in this method is to select an appropriate number of histogram bins. This directly impacts the quality of the resulting distribution vectors. If the bins are too narrow, the histogram becomes noisy and unstable, capturing random fluctuations rather than meaningful structure. On the other hand, overly wide bins can excessively smooth the distribution, erasing valuable information about the microstructural variability. To balance these effects, we employ three well-established bin width estimation strategies, each suited for different types of data.\\

Doane rule \cite{Doane1976} is applied to skewed or non-normal distributions. It is an improved version of Sturges’ formula that produces better estimates for non-normal data. We use it for Grain Disorientation, Grain Coordination Number, and the Grain Euler Angles $(\phi_1, \Phi, \phi_2)$. Rice rule \cite{RiceRule2024} offers a simple estimate based on data size alone and tends to slightly overestimate bin counts. We use it for Grain Size and Perimeter. Square Root rule \cite{MATLABHistogramDoc}, commonly used for its simplicity and speed, estimates the number of bins as the square root of the sample size. These adaptive strategies ensure that the resulting metric vectors reflect meaningful patterns in the data while avoiding overfitting to noise or oversmoothing relevant variations.

\subsection{Data Encoding}
\label{dataencoding}

\begin{figure}[H]
    \centering
    \includegraphics[width=0.9\textwidth]{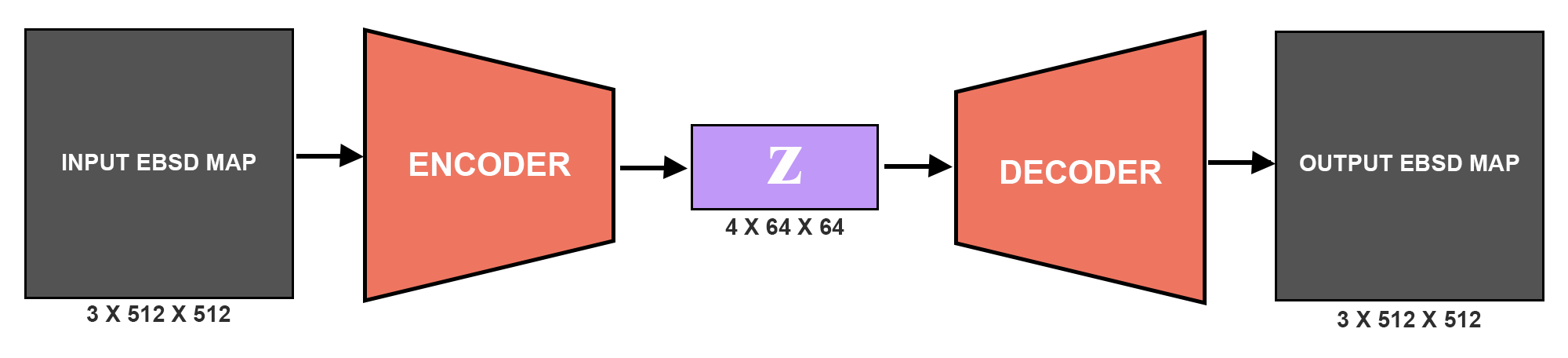}
    \caption{Overview of the VAE workflow. The input EBSD map is compressed by the Encoder to a latent space $Z$, and then reconstructed by the Decoder.}
    \label{fig:vae}
\end{figure}

As illustrated in Figure \ref{fig:vae}, we trained a VAE from Diffusers \cite{VonPlaten2022} to compress and reconstruct EBSD maps before and after applying the diffusion process. The VAE allows us to operate in a compressed representation of the EBSD map instead of the full-resolution data. Running the diffusion process in latent space drastically reduces computational cost and training time. The VAE learns to encode important semantic and structural features while removing noise and redundant details, which helps the generative model to generalize better by focusing only on meaningful compressed information.\\

Our VAE consists of an Encoder that transforms the input EBSD map $\mathbf{X} \in \mathbb{R}^{3 \times 512 \times 512}$ into a latent representation $\mathbf{Z} \in \mathbb{R}^{4 \times 64 \times 64}$, and a Decoder that reconstructs the original map from this representation. The overall structure is illustrated in Figure \ref{fig:vae}.\\

\subsection{Diffusion Network}
\label{diffusionnetwork}

Extending an EBSD map from a partial observation requires a model that can synthesize the missing regions while preserving the consistency of the known areas. To achieve this, we adopt a masking-based conditioning strategy similar to the one used in image inpainting with Stable Diffusion \cite{Rombach2022, VonPlaten2022}. In Stable Diffusion, inpainting is achieved by providing the model with a version of the image where missing regions are explicitly masked and conditioning the generative process on the visible part. This allows the model to generate content that is spatially coherent with the known regions, while filling in the missing parts in a semantically plausible way. This mechanism is particularly suited for EBSD map extension, where local orientation relationships and spatial continuity must be maintained across the generated regions. By feeding the model the partially masked EBSD map, along with a binary mask indicating the missing areas, the network can effectively reconstruct large, consistent EBSD maps.\\

To model the RP at the core of our diffusion-based approach, we use a UNet architecture \cite{Ronneberger2015} adapted from Diffusers \cite{VonPlaten2022}. The network is trained to predict the noise added to a latent representation of an EBSD map at a given diffusion timestep. Rather than operating directly on the EBSD map, the model performs all computations in the latent domain, which allows for a more compact representation and computational efficiency.\\

\begin{figure}[H]
    \centering
    \includegraphics[width=0.5\textwidth]{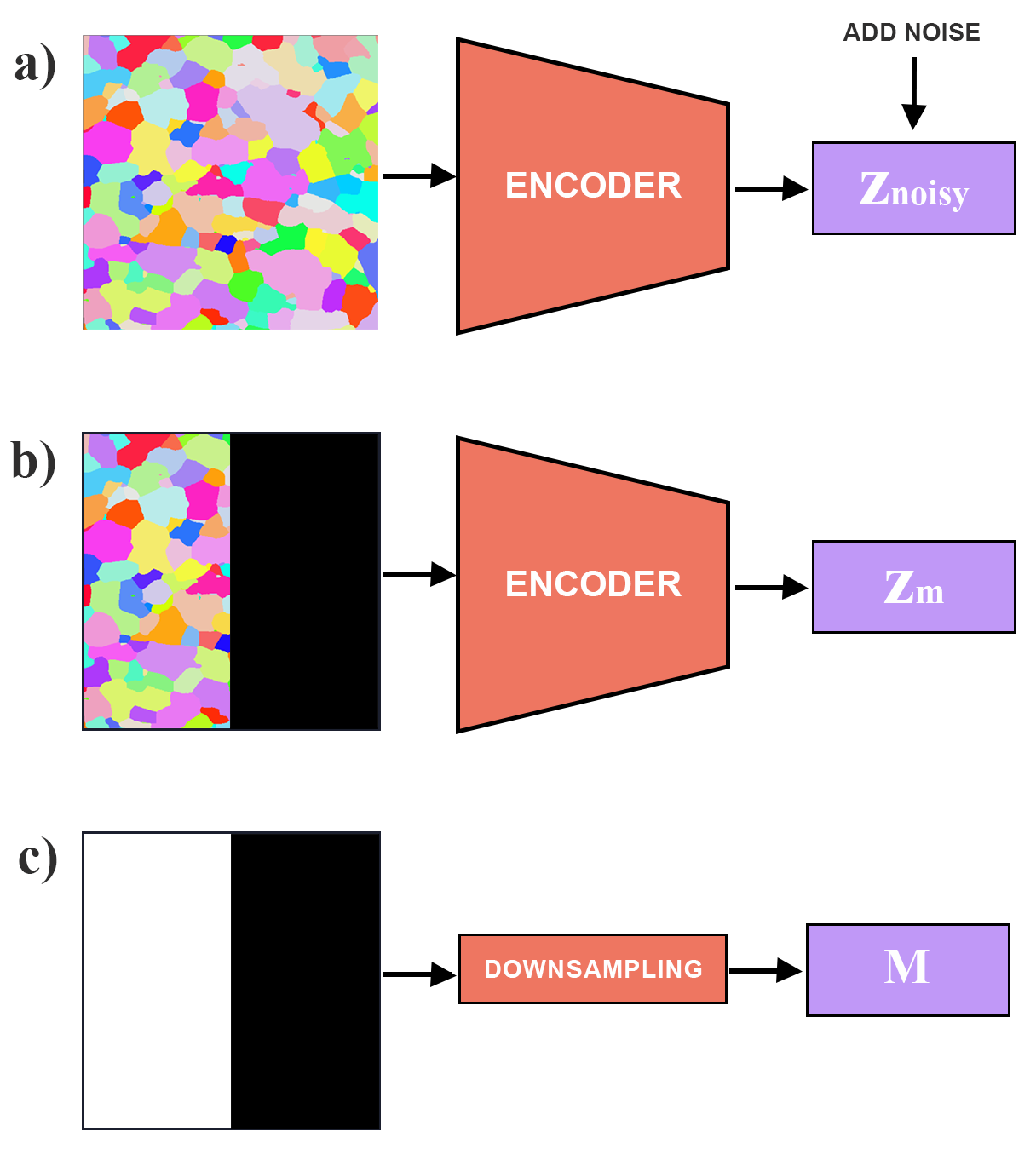}
    \caption{
        Latent encoding of the inputs used in the diffusion model. 
        \textbf{a)} The full EBSD map $\mathbf{X}$ is encoded by the VAE encoder and perturbed with Gaussian noise to produce the noisy latent $\mathbf{Z_{\text{noisy}}}$. 
        \textbf{b)} The partial (masked) EBSD map $\mathbf{X_m}$, corresponding to the known visible region, is also encoded into its own latent representation $\mathbf{Z_m}$. 
        \textbf{c)} The binary mask $\mathbf{M}$ is downsampled to match the latent resolution.
    }
    \label{fig:dataencoding}
\end{figure}

The encoding of input maps into this latent space is handled by the encoder component of the VAE, introduced in the previous section \ref{dataencoding}. As illustrated in Figure \ref{fig:dataencoding}, the encoder is used twice. It encodes the full EBSD input map $\mathbf{X} \in \mathbb{R}^{3 \times 512 \times 512}$ into a latent $\mathbf{Z} \in \mathbb{R}^{4 \times 64 \times 64}$, which is then perturbed with noise according to a predefined schedule at a diffusion timestep $\mathbf{t}$, producing the noisy latent $\mathbf{Z_{\text{noisy}}} \in \mathbb{R}^{4 \times 64 \times 64}$. Simultaneously, the partial EBSD map $\mathbf{X_m} \in \mathbb{R}^{3 \times 512 \times 512}$, corresponding to the unmasked region of the full input EBSD map $\mathbf{X}$ that needs to be extended, is encoded into its own latent representation $\mathbf{Z_m} \in \mathbb{R}^{4 \times 64 \times 64}$. In addition to these latents, we also prepare a binary mask $\mathbf{M} \in \mathbb{R}^{1 \times 512 \times 512}$ that identifies which parts of the input EBSD map are known (white) and which parts are missing and must be generated (black). This mask is downsampled to the latent resolution via bilinear interpolation \cite{Gonzalez2006}, resulting in a tensor $\mathbf{M} \in \mathbb{R}^{1 \times 64 \times 64}$. The timestep $t$ is embedded into a vector of size 1280, which is then used to condition the UNet across multiple layers.\\

\begin{figure}[H]
    \centering
    \includegraphics[width=0.9\textwidth]{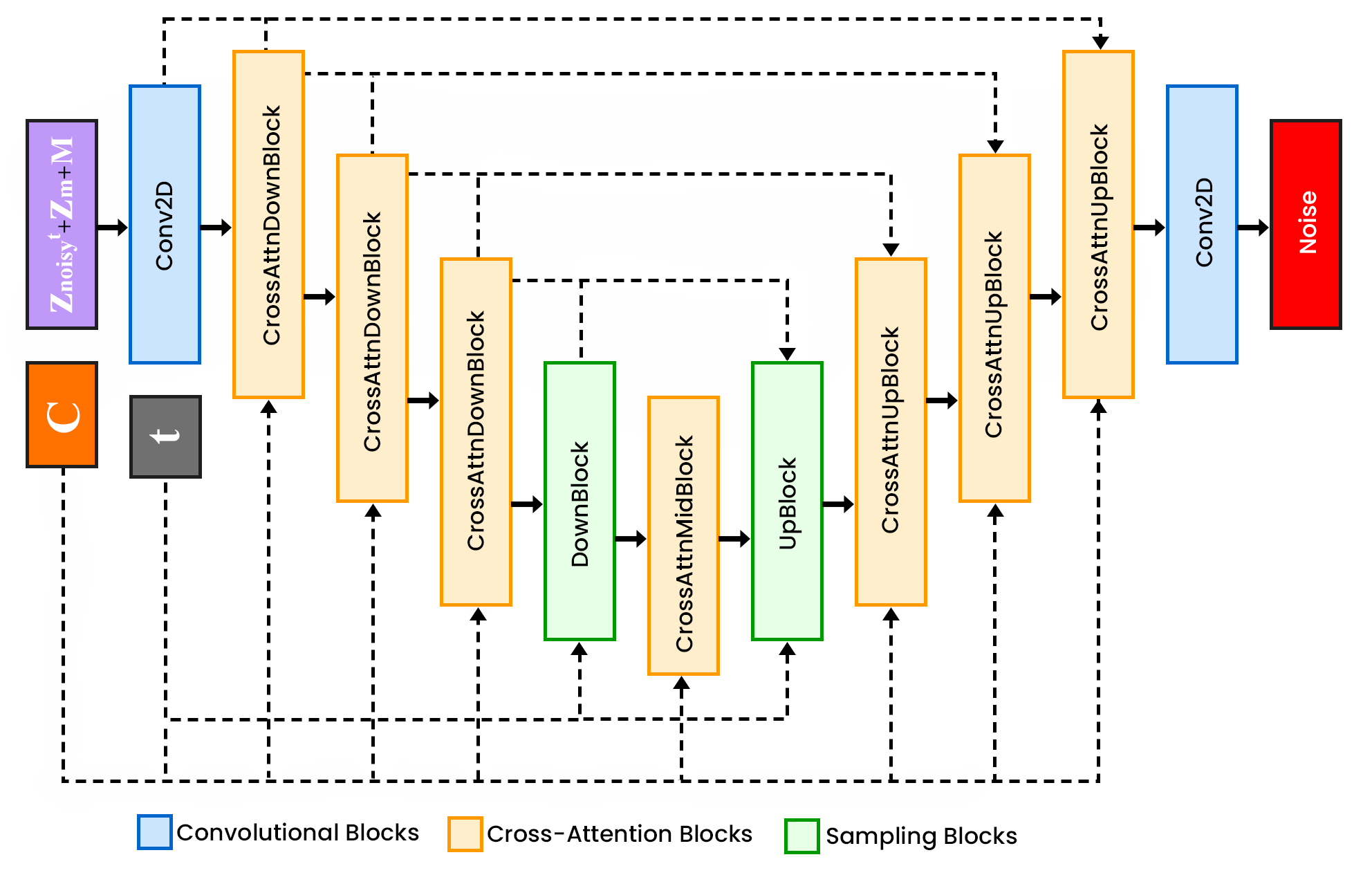}
    \caption{An overview of the UNet architecture.}
    \label{fig:unet}
\end{figure}

As shown in Figure \ref{fig:unet}, all three tensors, the noisy latent $\mathbf{Z_{\text{noisy}}}$, the masked latent $\mathbf{Z_m}$, and the binary mask $\mathbf{M}$ are concatenated along the channel dimension into a single input tensor $\mathbf{I} \in \mathbb{R}^{9 \times 64 \times 64}$. This composite representation is passed to the UNet, which also receives the scalar timestep $\mathbf{t}$ as an additional conditioning signal.\\ 

The UNet is further conditioned on the microstructural metrics described in Section \ref{dataprocessingmetrics}, which are encoded into a conditioning tensor $\mathbf{C} \in \mathbb{R}^{8 \times 1024}$. Rather than injecting this information directly as additional input channels, we leverage a cross-attention mechanism, described in Section \ref{attention}, to integrate these metrics into the UNet. Cross-attention provides a flexible and learnable way to modulate the internal activations of the network using an external conditioning signal. It enables each spatial location of the UNet's feature maps to selectively attend to relevant portions of the metric tensor $\mathbf{C}$, depending on the current stage of the denoising process and the characteristics of the partial input. For instance, when reconstructing a region located near a grain boundary, the model may need to pay more attention to the disorientation distribution or the coordination number, since these metrics reflect how grains interact and transition at their boundaries. In contrast, for areas located well inside a grain where the orientation tends to be more homogeneous the model might rely more on metrics like the inertia ratio and the Euler angle, which describe the internal geometry and crystallographic orientation of grains. The attention mechanism allows the network to adjust its predictions to improve both local realism and global consistency with the expected distribution of microstructural features.\\

The UNet, depicted in Figure \ref{fig:unet} and Table \ref{tab:unetblocks}, is responsible for predicting the noise $\mathbf{{\epsilon}_t} \in \mathbb{R}^{4 \times 64 \times 64}$ that was added to the clean latent $\mathbf{Z}$ during the FP. This predicted noise $\mathbf{\hat{\epsilon}_t}$ is then used to iteratively denoise $\mathbf{Z_{\text{noisy}}}$ and guide it toward a clean reconstruction of the underlying latent. By conditioning the UNet on both the partial EBSD map and the microstructural metrics through cross-attention, the model generates extended EBSD maps that are not only spatially consistent with the input but also statistically aligned with the desired microstructural properties. And once the final latent is recovered, it is decoded through the VAE decoder to generate the extended EBSD map for the partial map $\mathbf{X_m}$.\\

\begin{table}[H]
\centering
\resizebox{\textwidth}{!}{%
\begin{tabular}{lcc}
\toprule
\textbf{Block} & \textbf{Input} & \textbf{Output} \\
\midrule
Conv2D & (B, 9, 64, 64) & (B, 320, 64, 64) \\
CrossAttnDownBlock & (B, 320, 64, 64), (B, 1280), (B, 8, 1024) & (2x(B, 320, 32, 32)), (2x(B, 320, 64, 64)) \\
CrossAttnDownBlock & (B, 320, 32, 32), (B, 1280), (B, 8, 1024) & (2x(B, 640, 16, 16)), (2x(B, 640, 32, 32)) \\
CrossAttnDownBlock & (B, 640, 16, 16), (B, 1280), (B, 8, 1024) & (2x(B, 1280, 8, 8)), (2x(B, 1280, 16, 16)) \\
DownBlock & (B, 1280, 8, 8), (B, 1280) & (2x(B, 1280, 8, 8)) \\
CrossAttnMidBlock & (B, 1280, 8, 8), (B, 1280), (B, 8, 1024) & (B, 1280, 8, 8) \\
UpBlock & (4x(B, 1280, 8, 8)), (B, 1280) & (B, 1280, 16, 16) \\
CrossAttnUpBlock & (3x(B, 1280, 16, 16)), (B, 640, 16, 16), (B, 1280), (B, 8, 1024) & (B, 1280, 32, 32) \\
CrossAttnUpBlock & (B, 1280, 32, 32), (2x(B, 640, 32, 32)), (B, 320, 32, 32), (B, 1280), (B, 8, 1024) & (B, 640, 64, 64) \\
CrossAttnUpBlock & (B, 640, 64, 64), (3x(B, 320, 64, 64)), (B, 1280), (B, 8, 1024) & (B, 320, 64, 64) \\
Conv2D & (B, 320, 64, 64) & (B, 4, 64, 64) \\
\bottomrule
\end{tabular}
} % end resizebox
\caption{Summary of main blocks in the UNet architecture, with input and output tensor shapes at each stage. B refers to the batch size.}
\label{tab:unetblocks}
\end{table}

\subsection{Scheduler}
\label{scheduler}

In diffusion models, the scheduler defines the noise schedule by controling the noise level at each timestep during the diffusion processes \cite{Ho2020, Chen2023, Hang2024, Lin2024}. It controls how noise is added to clean inputs to simulate the FP, and guides how the model should iteratively remove noise at the RP. The scheduler is the computational engine that controls how noise is injected and removed, ensuring that the stochastic process remains mathematically coherent and practically efficient \cite{Sahoo2025, Sabour2024}. Choosing the appropriate scheduler is crucial for successful diffusion models \cite{Chang2026}.\\

In this work we use a Pseudo Numerical Methods for Diffusion Models (PNDM) scheduler \cite{Liu2022} to control the noise level at each timestep during the diffusion processes. This scheduler enables high-quality generation even with a reduced number of inference steps. Our implementation follows the design provided in the Diffusers \cite{VonPlaten2022}. We adopt a total of 50 diffusion steps $\left(\mathbf{T = 50}\right)$. 

\subsection{Training}
\label{training}

\subsubsection{Timestep Sampling}
\label{trainingtimestepsampling}

To ensure that the model learns to denoise effectively across the entire diffusion trajectory, we employ a random timestep sampling strategy. At each training iteration, a diffusion timestep $\mathbf{t}$ is randomly sampled from a uniform distribution over the total number of diffusion steps. This approach exposes the network to latents with varying noise levels from nearly pure noise to almost clean representations, helping the model generalize across the full denoising spectrum. As a result, the trained network becomes capable of handling the entire RP during inference, progressively refining pure noises into realistic EBSD reconstructions.

\subsubsection{Masks}
\label{trainingsmasks}

\begin{figure}[H]
    \centering
    \includegraphics[width=0.9\textwidth]{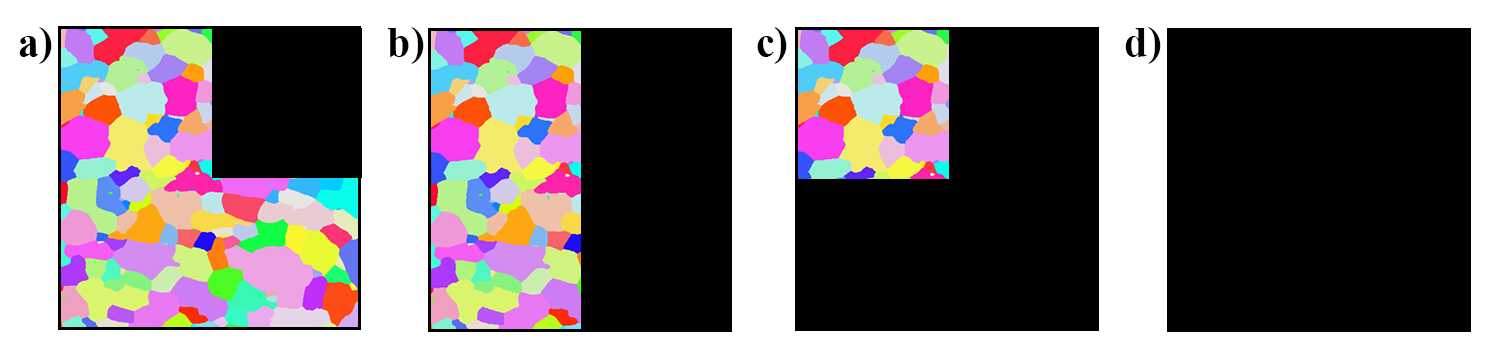}
    \caption{
        The four types of binary masks used during training. a) Closed-angle Mask: the model learns to extend diagonally across two adjacent sides. b) Horizontal and Vertical Extension Mask: used to train the model for lateral or vertical continuation. c) Open-angle Mask: only a corner region is provided. d) Full Mask: no EBSD input is provided, allowing generation solely from microstructural metrics.
    }
    \label{fig:masks}
\end{figure}

To train the model to handle different spatial configurations of missing data, we designed four types of binary masks, illustrated in Figure \ref{fig:masks}. Each mask defines a specific geometric configuration of the visible and missing regions in the EBSD map, enabling the diffusion model to learn how to infer microstructural continuity under diverse spatial constraints.

Mask a) corresponds to an closed-angle configuration, where two adjacent sides of the map are visible while the opposite corner remains missing. This pattern trains the model to synthesize new regions extending diagonally. Mask b) defines an horizontal and vertical extension configuration, in which half of the map is missing along one direction. This setup encourages the model to learn how to grow microstructures laterally or vertically. Mask c) represents an open-angle configuration, where only a corner portion of the map is visible. This case is particularly challenging, as it requires the model to infer large missing areas based on minimal spatial context. Finally, Mask d) corresponds to the scenario where no input EBSD map is provided. In this case the model must generate a complete EBSD map solely from the conditioning metrics, without any spatial guidance.\\

During training, for each sample within a batch, one of the four mask types is randomly selected and applied to the input map. This random sampling ensures that the model is exposed to a wide range of reconstruction scenarios, improving its generalization ability. The practical use of these masks within our patch-wise geometric extension strategy for generating EBSD maps of arbitrary size is discussed in Section \ref{ases}.

\subsubsection{Loss Function}
\label{traininglossfunction}

The predicted noise $\mathbf{\hat{\epsilon}_t}$ is compared to the ground-truth noise $\mathbf{\epsilon}_t$ using a Mean Squared Error (MSE) loss function \cite{Harville1992, Goodfellow2016}. This loss function measures how accurately the model estimates the stochastic perturbation introduced at each step of the FP, and is defined as:

\begin{equation}
\mathcal{L}_{\text{MSE}} = \frac{1}{\mathbf{N}} \sum_{i=1}^{\mathbf{N}} (\mathbf{\epsilon}_ti - \mathbf{\hat{\epsilon}}_ti)^2, \quad \text{with} \quad \mathbf{N} = \mathbf{B} \times \mathbf{C} \times \mathbf{H} \times \mathbf{W}
\end{equation}

Here, $\mathbf{B}$ denotes the training batch size, $\mathbf{C}$ the number of latent channels, and $\mathbf{H}$ and $\mathbf{W}$ the spatial dimensions of the latent tensor.

\subsubsection{Optimization}
\label{trainingoptimization}

The network parameters are optimized using the AdamW optimizer \cite{Loshchilov2017, Paszke2019}, a variant of the Adam algorithm \cite{Kingma2014Adam} that decouples weight decay from the gradient update. This separation helps maintain a more stable optimization trajectory, particularly for deep convolutional architectures like the UNet used in our diffusion model. The optimization objective aims to minimize the MSE loss described in Section \ref{traininglossfunction}, ensuring accurate noise prediction across all diffusion timesteps.\\

To control the learning dynamics during training, the learning rate is scheduled following a Cosine Annealing strategy \cite{Loshchilov2016, He2019, Senior2013}. This schedule gradually decreases the learning rate following a half-cosine curve, allowing large updates at the beginning of training and smaller refinements as convergence is approached. The learning rate at each epoch is defined by:

\begin{equation}
\eta_t = \eta_{\text{min}} + \frac{1}{2}(\eta_{\text{max}} - \eta_{\text{min}})\left(1 + \cos\left(\frac{T_{\text{cur}}}{T_{\text{max}}} \pi\right)\right),
\end{equation}

where $\eta_{\text{min}}$ and $\eta_{\text{max}}$ denote the minimum and maximum learning rates, $\mathbf{T_{\text{cur}}}$ is the current training epoch, and $\mathbf{T_{\text{max}}}$ is the total number of training epochs. This smooth decay mechanism prevents abrupt changes in the optimization process and helps avoid oscillations near convergence, while maintaining efficient exploration. 

\subsection{Inference Process}
\label{inferenceprocess}

\begin{figure}[H]
    \centering
    \includegraphics[width=0.9\textwidth]{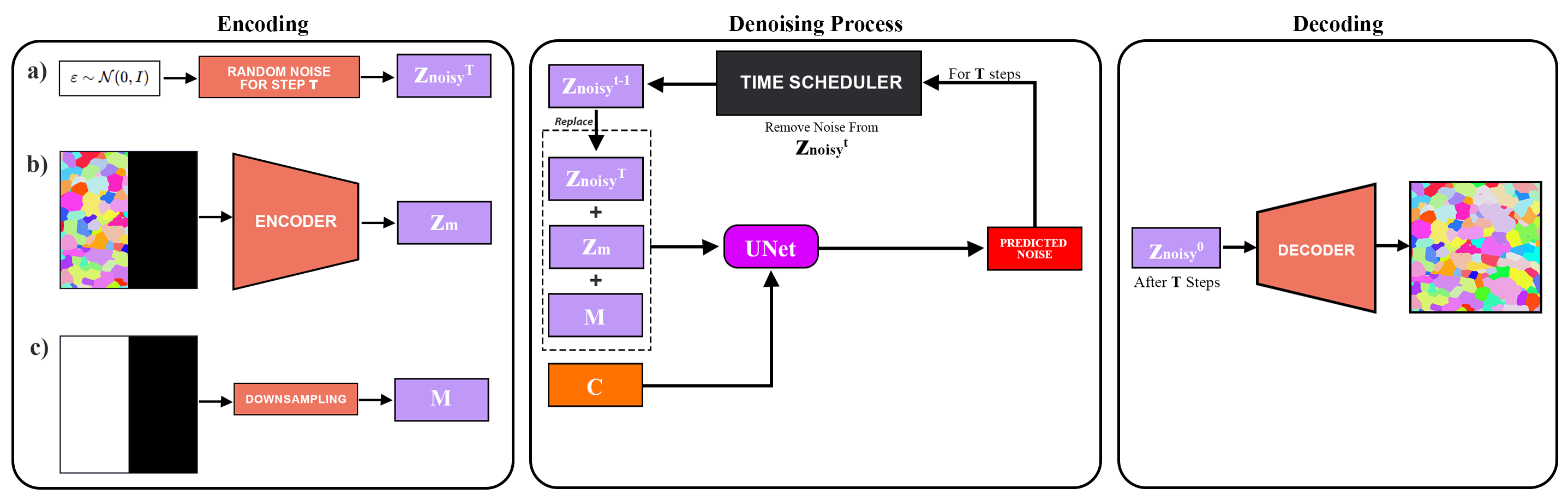}
    \caption{
        Overview of the inference process. Encoding phase, where (a) Gaussian noise is sampled to create $\mathbf{Z}_{\text{noisy}}^T$, (b) a partial EBSD map is encoded into $\mathbf{Z_m}$, and (c) a binary conditioning mask $\mathbf{M}$. During denoising, the scheduler and UNet iteratively predict and remove noise from $\mathbf{Z}_{\text{noisy}}^t$, guided by the microstructural metrics vector $\mathbf{C}$. The final latent $\mathbf{Z}_{\text{noisy}}^0$ is decoded to produce the complete EBSD map.}
    \label{fig:inferenceprocess}
\end{figure}

As illustrated in Figure \ref{fig:inferenceprocess}, the inference begins with a Gaussian noise $\boldsymbol{\epsilon} \sim \mathcal{N}(0, \mathbf{I})$ sampled to produce an initial noisy latent $\mathbf{Z}_{\text{noisy}}^T$. A partial EBSD map is encoded by the VAE encoder into its latent representation $\mathbf{Z_m}$. In parallel, the corresponding binary mask $\mathbf{M}$ is downsampled to match the latent resolution.

We concatenated the current noisy latent $\mathbf{Z}_{\text{noisy}}^t$, the masked latent $\mathbf{Z_m}$, and the binary mask $\mathbf{M}$ along the channel dimension as input for the UNet. It is also conditioned on the metrics tensor $\mathbf{C}$. Through cross-attention, these metrics guide the model toward generating physically consistent structures that respect the expected statistical properties. Given this input, the UNet predicts the noise component $\hat{\boldsymbol{\epsilon}}_t$ present in $\mathbf{Z}_{\text{noisy}}^t$. The scheduler then uses this prediction to update the latent and compute $\mathbf{Z}_{\text{noisy}}^{t-1}$, effectively removing part of the noise. The denoising loop operates iteratively over $T$ timesteps, starting from $\mathbf{Z}_{\text{noisy}}^T$ and progressively refining it into a clean latent $\mathbf{Z}_{\text{noisy}}^0$.\\ 

Once the denoising loop reaches the final iteration, the resulting latent $\mathbf{Z}_{\text{noisy}}^0$ represents the clean, fully reconstructed EBSD latent. This tensor is then decoded back using the VAE decoder, recovering the complete EBSD map. The final reconstruction merges the initially observed region with the newly generated one, yielding a physically plausible and spatially coherent microstructure.

\subsection{Arbitrary Size Extension Strategy}
\label{ases}

\begin{figure}[H]
    \centering
    \includegraphics[width=0.7\textwidth]{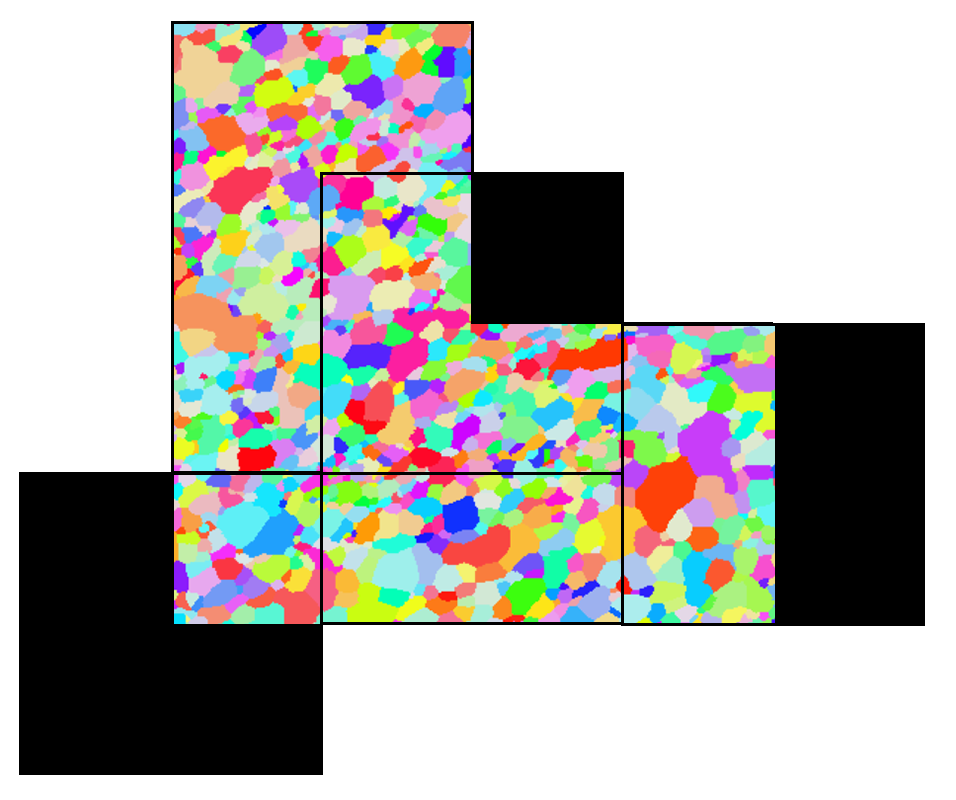}
    \caption{
        Illustration of the patch-wise geometric extension strategy used to synthesize large EBSD maps. Starting from an arbitrary size EBSD map input, the generation proceeds iteratively in multiple directions using different spatial masks.
    }
    \label{fig:arbitraryextension}
\end{figure}

The ability to generate EBSD maps of arbitrary size requires going beyond the standard fixed-size inference used during inference. For this task, we adopt a patch-wise extension strategy in which the model is repeatedly applied over overlapping windows, progressively synthesizing unseen regions based on partially visible contexts. This approach is illustrated in Figure \ref{fig:arbitraryextension}. Given an initial EBSD map we iteratively expand the known region by generating new surrounding patches using the trained diffusion model. Each new patch overlaps partially with the previously generated ones. Depending on its relative position to the known region, a specific mask configuration is selected from the four types defined in Section \ref{trainingsmasks}. These masks allow the model to adapt its generation strategy to the local context. During this extension process, each patch is generated by calling the inference pipeline described in Section \ref{inferenceprocess}.\\

As shown in Figure \ref{fig:graphical_abstract}, with this strategy, an EBSD map of arbitrary size and shape can be provided as input, with a minimum required size of $256\times256$ cells, along with a target output size. The model can therefore extend the input microstructure to the desired dimensions. Once the complete EBSD map has been generated, it is exported in the standardized .ctf file. A post-processing is then performed in MTEX, where the reconstructed crystallographic orientations are filtered using the same procedure described in Section \ref{dataprocessingebsd}. This additional filtering step eliminates residual noise from the generative process while preserving the geometric and crystallographic consistency of the extended microstructure.

\section{Experiments}
\label{experiments}

\subsection{Dataset}
\label{experimentsdataset}

Our experiments rely on a dataset of 70 EBSD maps, obtained from real experimental acquisitions, mostly provided by the CEMEF Laboratory. A part of the Inconel 625 data is coming from a joint project between several european laboratories as part of the 2023 ESAFORM Benchmark \cite{Agirre2025a, Agirre2025b}. All the microstructures considered are free of second phase particles and fully recrystallized. Each map was processed using the method described in Section \ref{dataprocessing}. The dataset covers four different metallic alloys, each selected for their distinct microstructural configurations and industrial significance. Table \ref{tab:dataset_summary} provides a summary of the materials included, the number of EBSD samples per material, and a brief description of their relevance. By including materials with different profiles, the model is exposed to a wide range of realistic metallurgical configurations, which improves its ability to synthesize physically plausible EBSD maps. After the cropping phase, the original 70 EBSD maps were expanded to a total of 500 smaller patches of $512\times512$ cells. Following the data augmentation procedure, the 500 patches were augmented to 4000, enhancing the model’s generalization capability during training.

\begin{table}[H]
\centering
\begin{tabular}{l c p{8cm}}
\textbf{Material} & \textbf{EBSD Maps} & \textbf{Description} \\
\hline
\textbf{304L} & 14 & Austenitic stainless steel widely used in structural and nuclear components due to its corrosion resistance and weldability. \\
\textbf{316L} & 17 & Low-carbon austenitic stainless steel with enhanced corrosion resistance, particularly in chloride environments; commonly used in marine, nuclear and biomedical applications. \\
\textbf{In625} & 14 & Nickel-based superalloy known for its high strength and oxidation resistance at elevated temperatures, often used in chemical processing and aerospace. \\
\textbf{In718} & 25 & Precipitation-hardened nickel superalloy with excellent mechanical properties and creep resistance, widely employed in jet engines and high-performance turbines. \\
\end{tabular}
\caption{Summary of materials and EBSD map counts used in the dataset.}
\label{tab:dataset_summary}
\end{table}

\subsection{Hyperparameters, Hardware and Timings}
\label{experimentshht}

All experiments were conducted using a single NVIDIA A100-PCIE-40GB GPU. The diffusion model was trained over 1500 epochs, with a batch size of 8 and an initial learning rate set to $1 \times 10^{-5}$. Each training epoch required approximately 10 minutes, resulting in a total training time of around 250 hours. During inference, the generation of a single $512\times512$ cells EBSD patch takes roughly 2.5 seconds. We also measured the total inference time required to extend maps at different target resolutions using our patch-wise extension strategy. Extending an initial $512\times512$ cells map to $1024\times1024$ requires approximatively 20 seconds, to $1536\times1536$ takes approximatively 60 seconds, to $2048\times2048$ requires approximatively 123 seconds and to a larger $3072\times3072$ cells map takes approximatively 315 seconds. These timings reflect the sequential nature of our current implementation, where patches are synthesized one after another as the algorithm get through the surfaces to be filled. In principle, this process could be significantly accelerated through parallelization, since multiple patches could be generated concurrently whenever they do not overlap spatially. This makes the method computationally accessible for high-resolution microstructure synthesis. Furthermore, it should be emphasized that direct Laguerre-Voronoï-based generation methods also entail a non-negligible computational cost when precise adherence to grain size distributions is required, owing to the dense sphere-packing algorithms that typically underpin these approaches \cite{Hitti2012,Ilin2016}.

\section{Results}
\label{results}

\subsection{Input EBSD Map Extension}
\label{resultsextension}

\begin{figure}[H]
    \centering
    \includegraphics[width=0.4\linewidth]{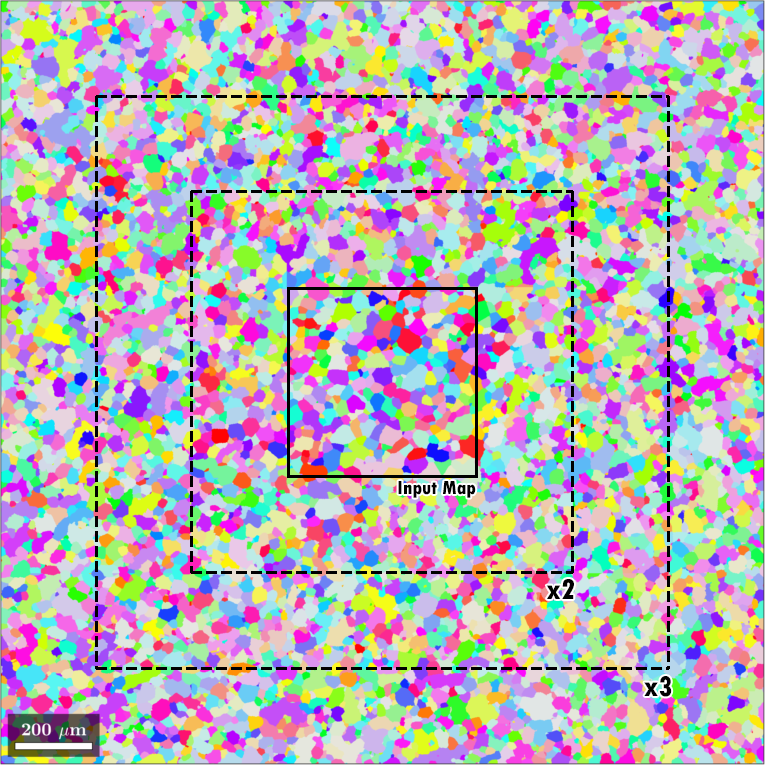}
    \includegraphics[width=0.4\linewidth]{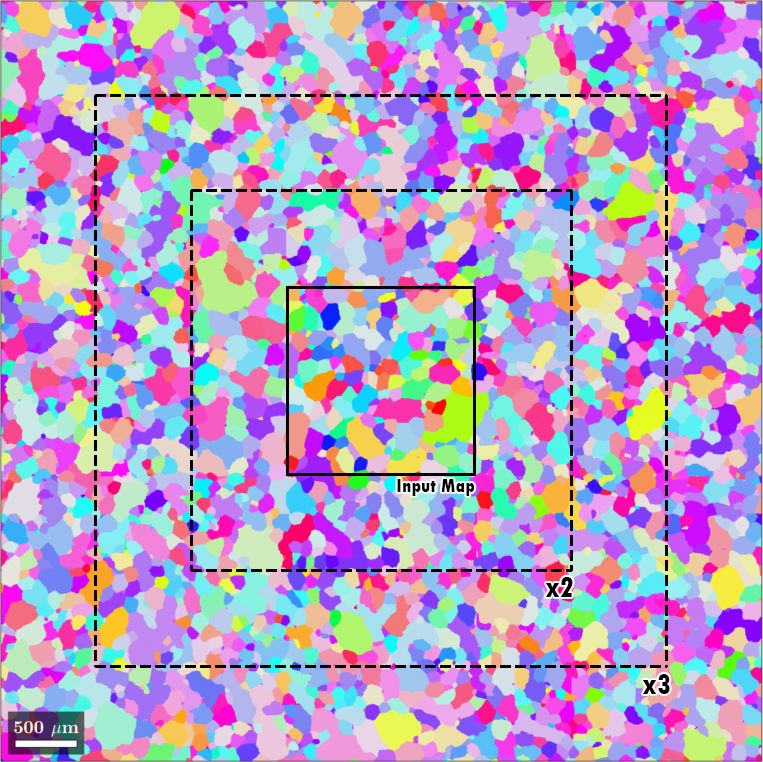}
    \includegraphics[width=0.1\linewidth]{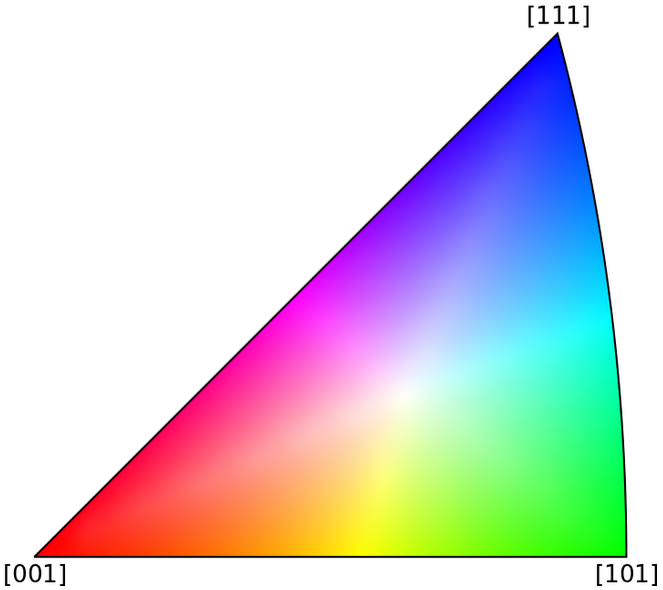}
    \caption{$2048\times2048$ Cells Extended EBSD Map of In718 (Left) and In625 (Right). The central $512\times512$ cells region corresponds to the input map. The surrounding regions are generated by the model while preserving morphological and crystallographic coherence. The Inverse Pole Figure (IPFZ) color key used for visualizing crystallographic orientations for all the generated EBSD Maps is also depicted within the standard orientation triangle.}
    \label{fig:extendedmaps}
\end{figure}

Figure \ref{fig:extendedmaps} illustrates two representative results from our method, showcasing the spatial coherence of the generated microstructures, one for In718 and another for In625. Each input map, of size $512\times512$ cells, is extended into a $1024\times1024$ cells map (x2), which is then extended to a $1536\times1536$ cells map (x3) and finally extended to a $2048\times2048$ cells map (x4). In both figures, the target microstructural metrics used to guide the generation come from the original input maps. For visualization purposes, all EBSD orientation maps presented in the following use the Inverse Pole Figure (IPFZ) color scheme as shown in the right side of the Figure \ref{fig:extendedmaps}.

Visually, the generated regions show a high degree of continuity in terms of grain morphology and orientation patterns. Grain boundaries are well preserved across the input borders, and there is no obvious tiling artifact. The model adapts the generation to the surrounding grains without requiring explicit boundary tracking. The extended maps include grains with morphologies that differ from those present in the original input region, suggesting that the model is not simply replicating local patterns but instead generating completely new statistically coherent structures aligned with the global conditioning metrics.

\begin{figure}[H]
    \centering
    \includegraphics[width=0.4\linewidth]{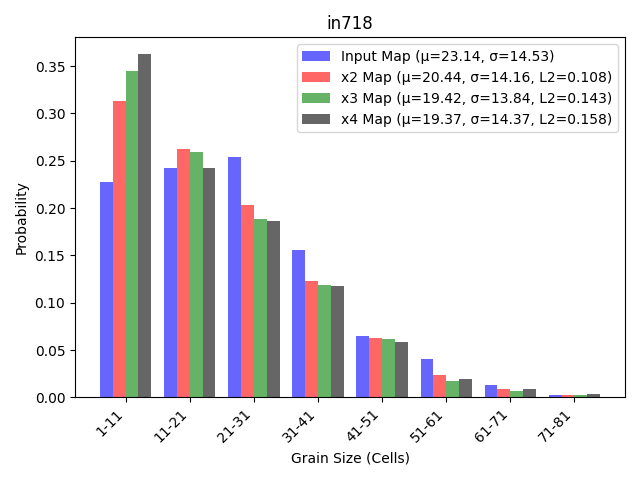}
    \includegraphics[width=0.4\linewidth]{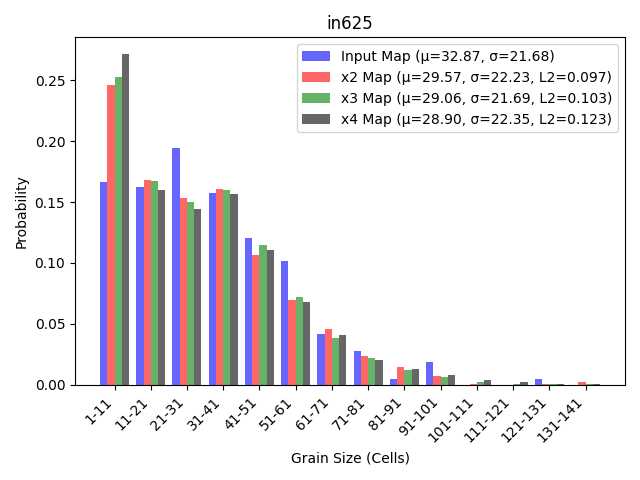}
    \caption{Comparison of Grain Size Distributions between the Input EBSD Map and the Extended EBSD Maps at different scales.}
    \label{fig:grainsize}
\end{figure}

\begin{figure}[H]
    \centering
    \includegraphics[width=0.4\linewidth]{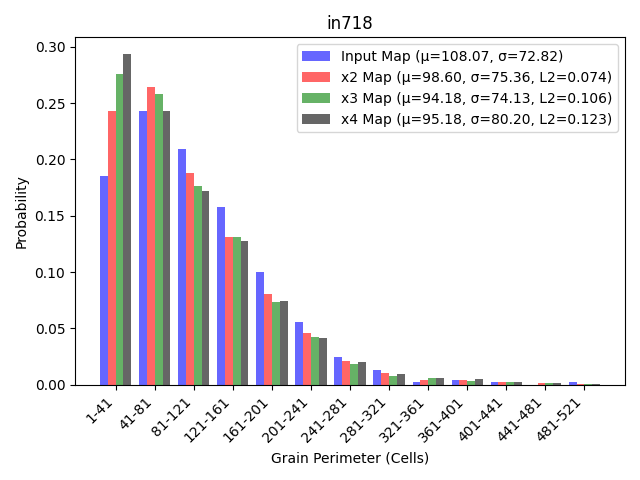}
    \includegraphics[width=0.4\linewidth]{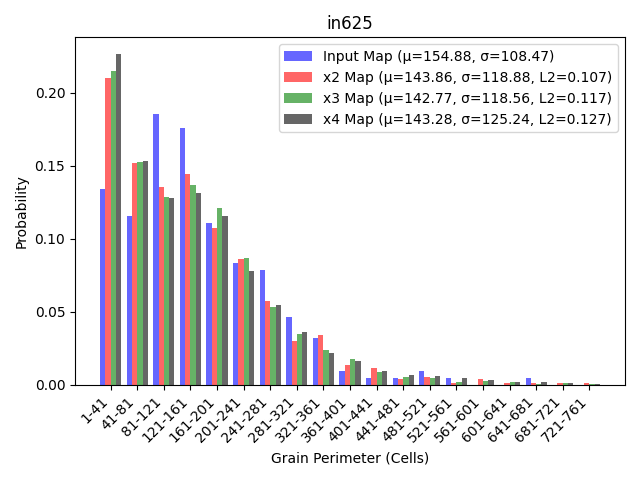}
    \caption{Comparison of Grain Perimeter Distributions between the Input EBSD Map and the Extended EBSD Maps at different scales.}
    \label{fig:grainperimeter}
\end{figure}

\begin{figure}[H]
    \centering
    \includegraphics[width=0.4\linewidth]{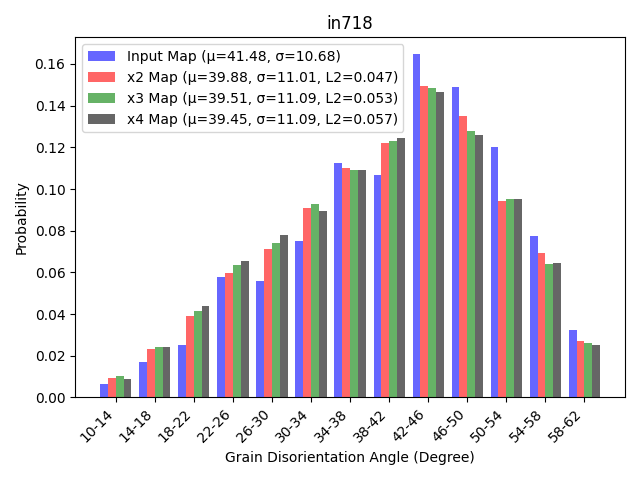}
    \includegraphics[width=0.4\linewidth]{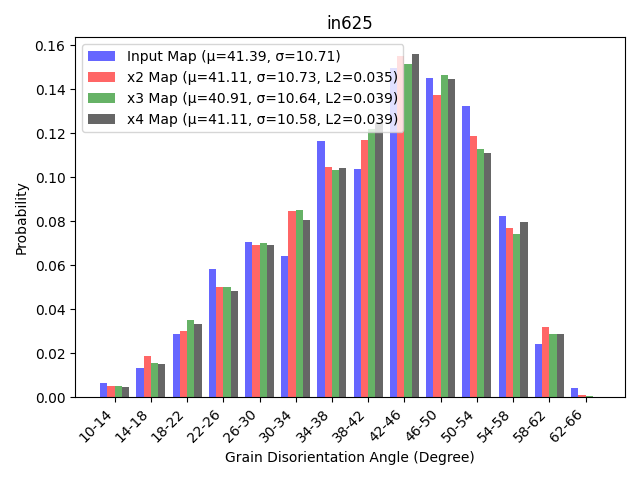}
    \caption{Comparison of Grain Boundary Disorientation Distributions between the Input EBSD Map and the Extended EBSD Maps at different scales.}
    \label{fig:grainmisorientation}
\end{figure}

\begin{figure}[H]
    \centering
    \includegraphics[width=0.4\linewidth]{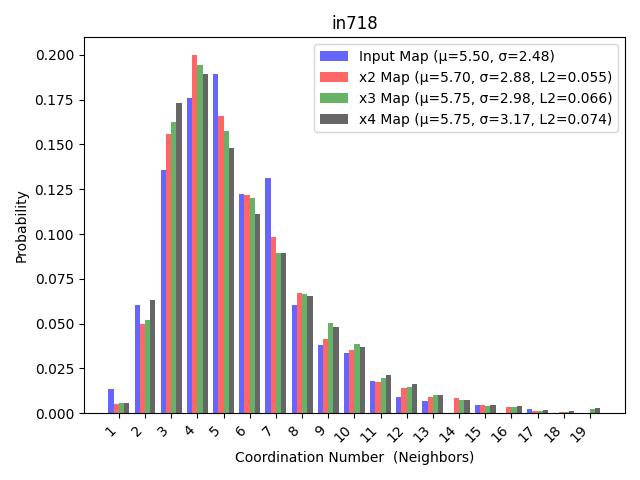}
    \includegraphics[width=0.4\linewidth]{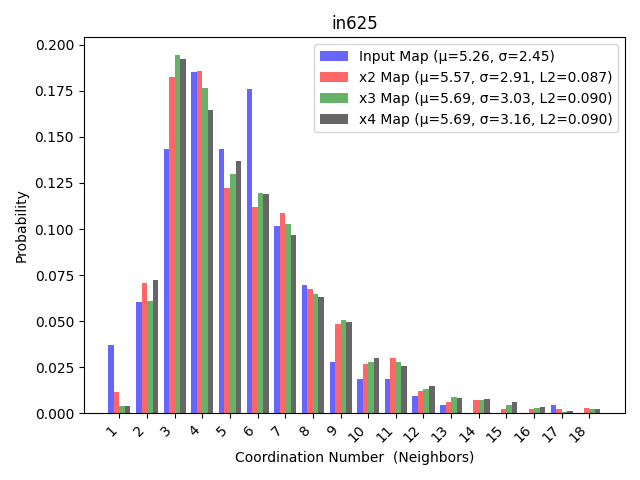}
    \caption{Comparison of Grain Coordination Number Distributions between the Input EBSD Map and the Extended EBSD Maps at different scales.}
    \label{fig:grainneighbors}
\end{figure}

\begin{figure}[H]
    \centering
    \includegraphics[width=0.4\linewidth]{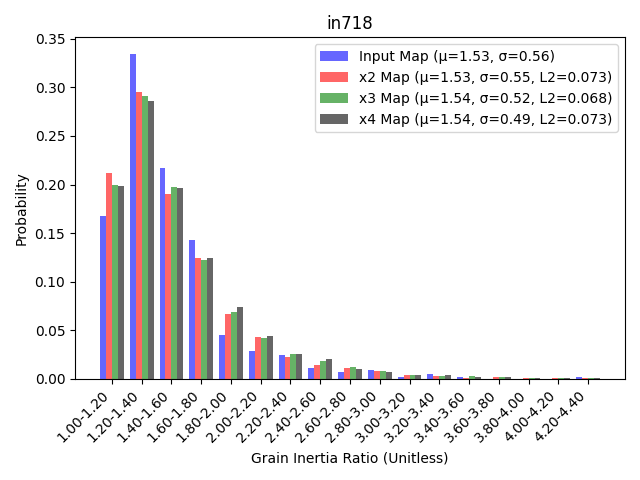}
    \includegraphics[width=0.4\linewidth]{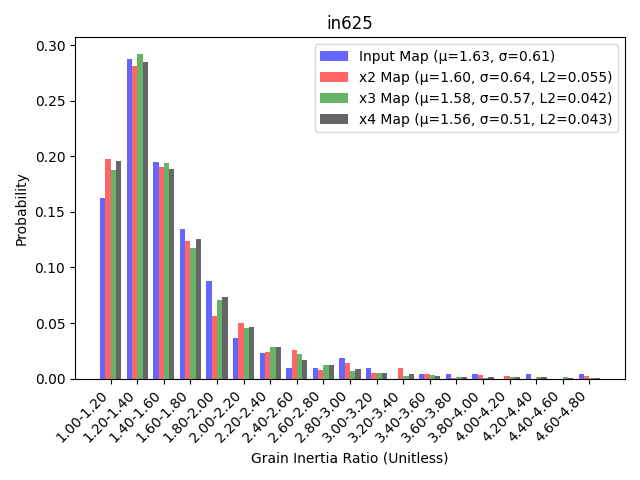}
    \caption{Comparison of Grain Inertia Ratio Distributions between the Input EBSD Map and the Extended EBSD Maps at different scales.}
    \label{fig:inertiaratio}
\end{figure}

To quantitatively assess the statistical plausibility of the generated microstructures, we compared the distributions of microstructural metrics between the conditioning input map $\mathbf{C}$ and the extended maps at different scales (x2, x3, x4). The comparisons are reported across the eight metrics described in Section \ref{dataprocessingmetrics}. Each Figures (\ref{fig:grainsize}, \ref{fig:grainperimeter}, \ref{fig:grainmisorientation}, \ref{fig:grainneighbors}, \ref{fig:inertiaratio}) shows the ground-truth distribution used to condition the generation (Input Map) and the metric computed on the extended maps at different scales. All metrics are computed using the same post-processing filters described in Section \ref{dataprocessingmetrics}. L2 is the norm L2 between the respective distributions of the extended maps and the input map.

\begin{figure}[H]
    \centering
    \includegraphics[width=0.4\linewidth]{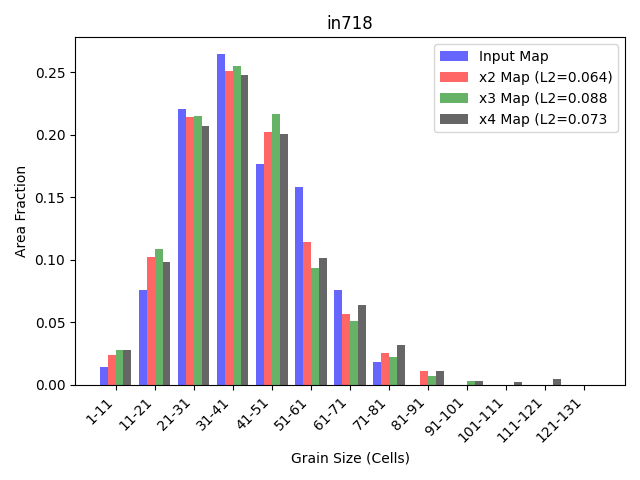}
    \includegraphics[width=0.4\linewidth]{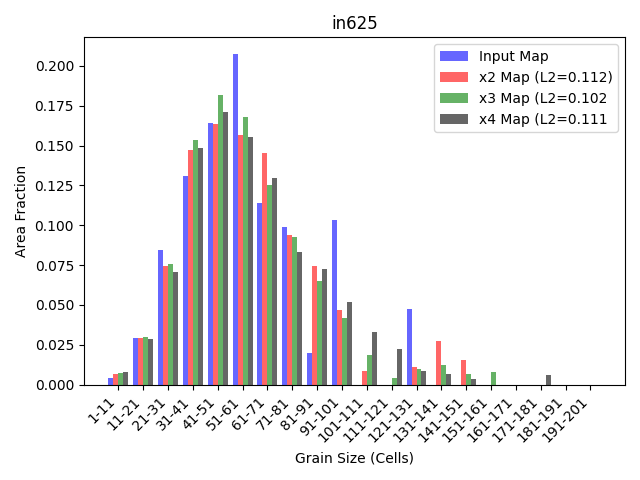}
    \caption{Comparison of area fraction per grain size bin at different extension scales. For each grain size bin, the proportion of the total map surface covered by grains within that size range is computed and compared across the Input EBSD Map and the Extended EBSD Maps at different scales.}
    \label{fig:areafraction}
\end{figure}

To complete the analysis, the area-weighted grain size distributions also depicted in Figure \ref{fig:areafraction} for the two maps presented in Figure \ref{fig:extendedmaps}. For each bin, we estimate the equivalent grain area as $\pi r^2$ for each grain, where $r$ is half the equivalent diameter. The areas are then summed and normalized by the total map area. 

\begin{figure}[H]
    \centering
    \includegraphics[width=0.6\linewidth]{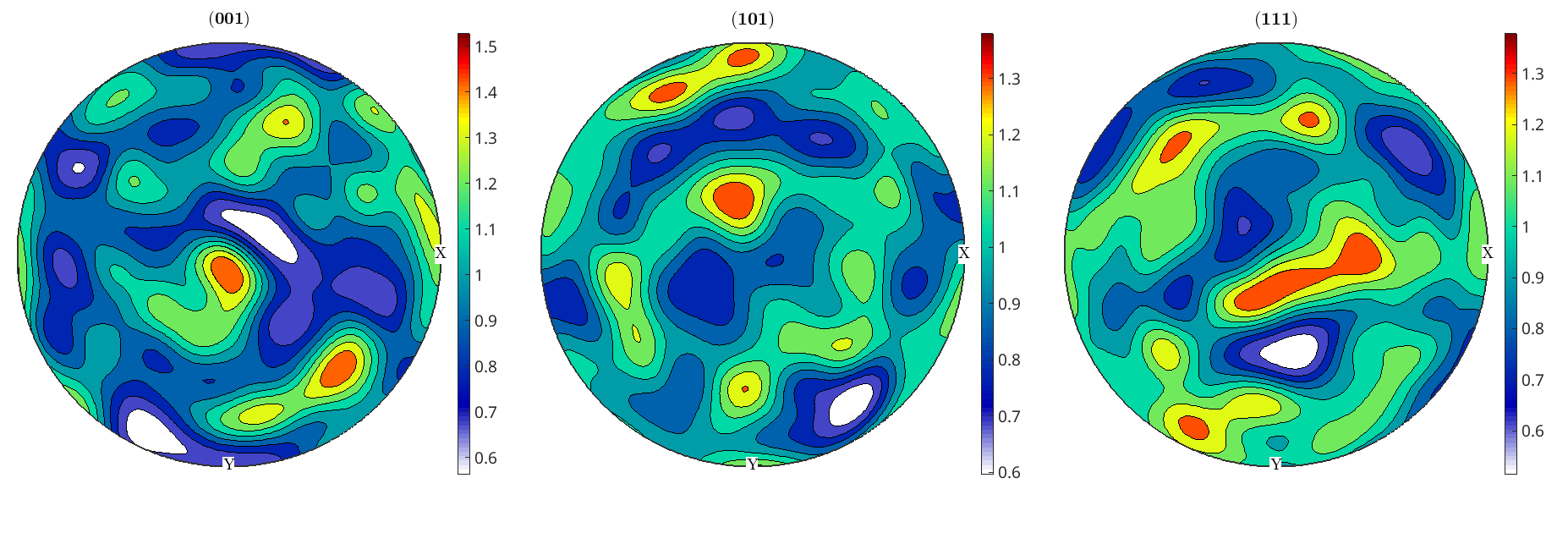}
    \includegraphics[width=0.6\linewidth]{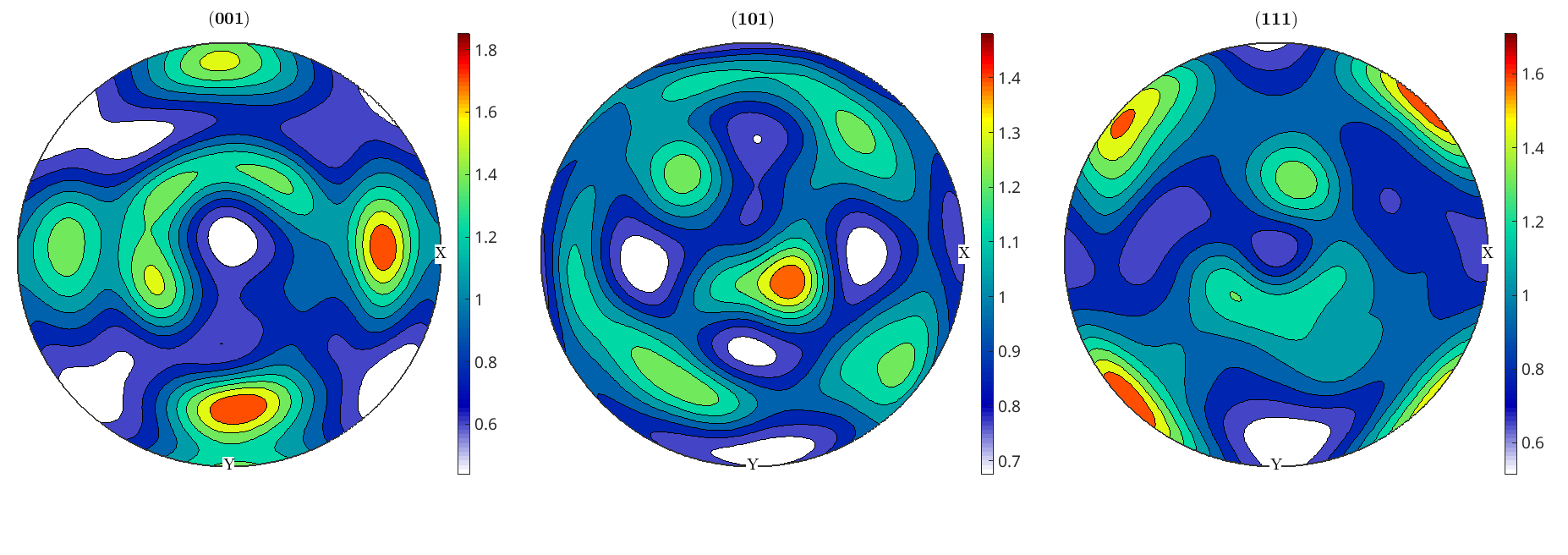}
    \caption{Comparison of Orientation Distribution Function (ODF) for the input $512\times512$ map (top) and the extended  $2048\times2048$ map (bottom) for In718 map of Figure \ref{fig:extendedmaps}.}
    \label{fig:dofin718}
\end{figure}

\begin{figure}[H]
    \centering
    \includegraphics[width=0.6\linewidth]{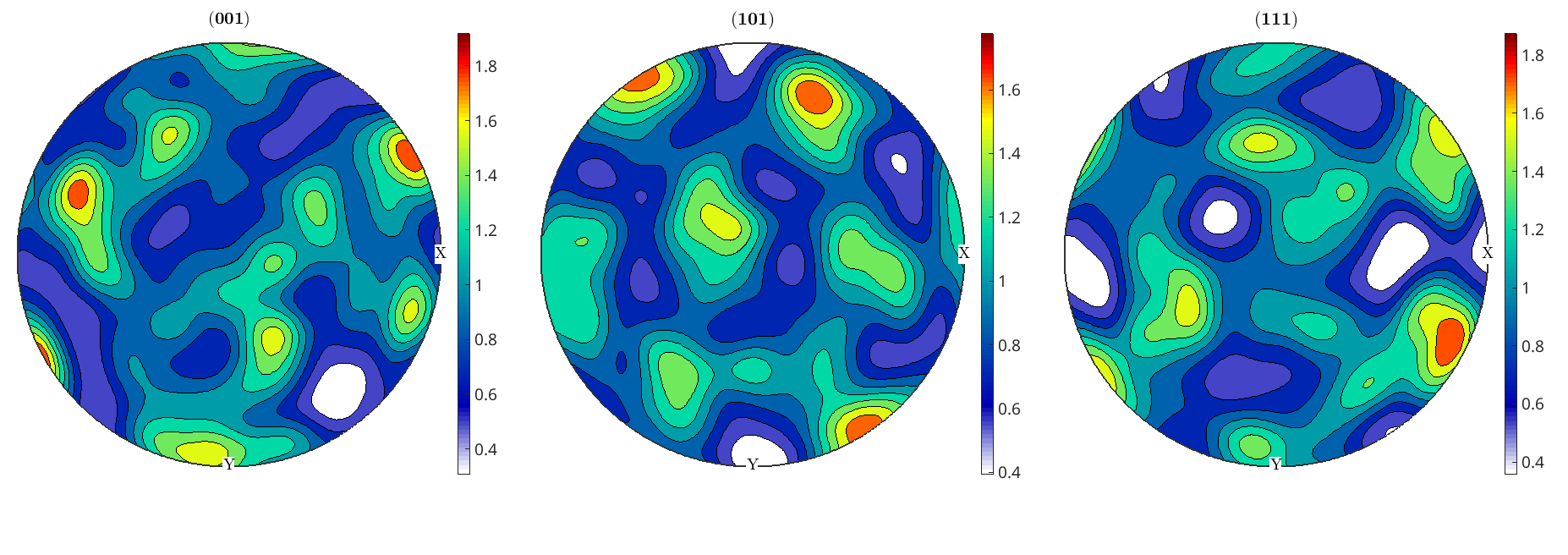}
    \includegraphics[width=0.6\linewidth]{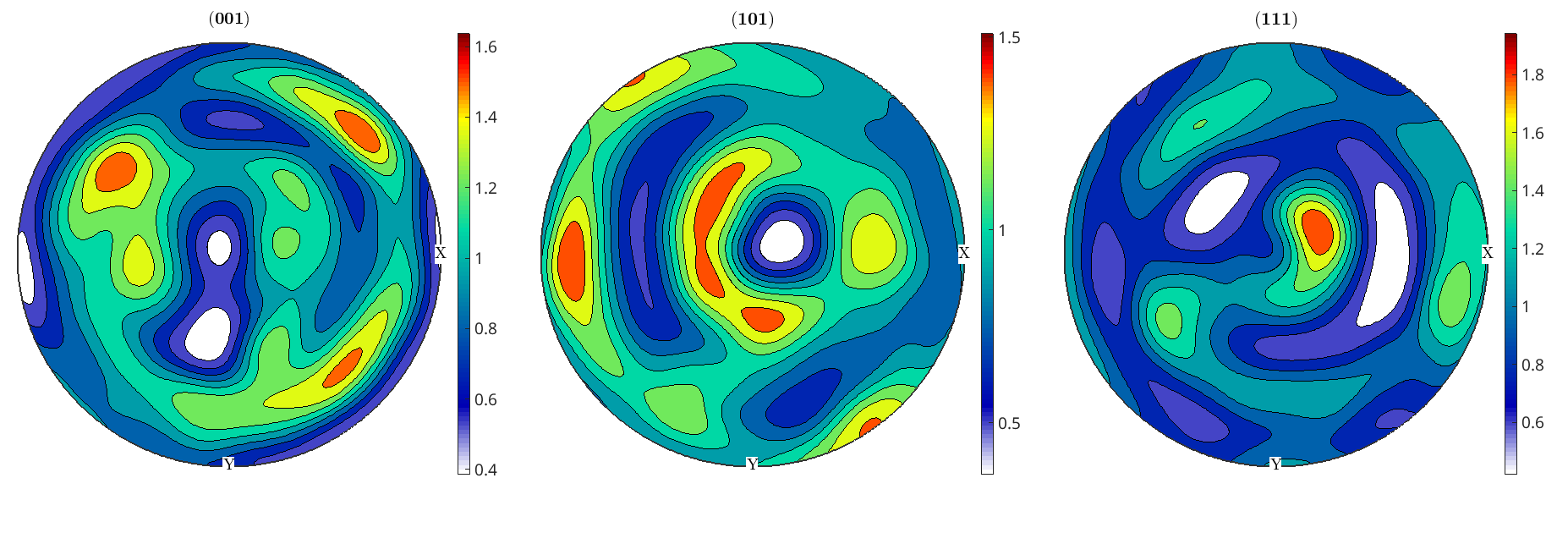}
    \caption{Comparison of Orientation Distribution Function (ODF) for the input $512\times512$ map (top) and the extended  $2048\times2048$ map (bottom) for In625 map of Figure \ref{fig:extendedmaps}.}
    \label{fig:dofin625}
\end{figure}

To further examine the crystallographic consistency of the generated microstructures, we compared the Orientation Distribution Functions (ODFs) of the input maps and their extended version (x4) shown in Figure \ref{fig:extendedmaps}. Since the EBSD maps used in this study correspond to materials with uniform orientation distributions, no pronounced texture is expected. As shown in Figures \ref{fig:dofin718} and \ref{fig:dofin625}, the ODF of the generated $2048\times2048$ maps exhibit patterns that remain consistent with those of the original $512\times512$ inputs, without the emergence of artificial texture components. This indicates that the diffusion model preserves the underlying randomness of the crystallographic orientations.

\subsection{Generated Maps (x4) Homogeneity}
\label{resultshomogeneity}

To assess the spatial homogeneity of the larger extended maps, the $2048\times2048$ maps (x4) (see Figure \ref{fig:extendedmaps}), we analyze whether local regions maintain statistical consistency with the entire microstructure. This is a crucial aspect when considering synthetic maps for downstream simulations. 

\begin{figure}[H]
    \centering
    \includegraphics[width=0.49\linewidth]{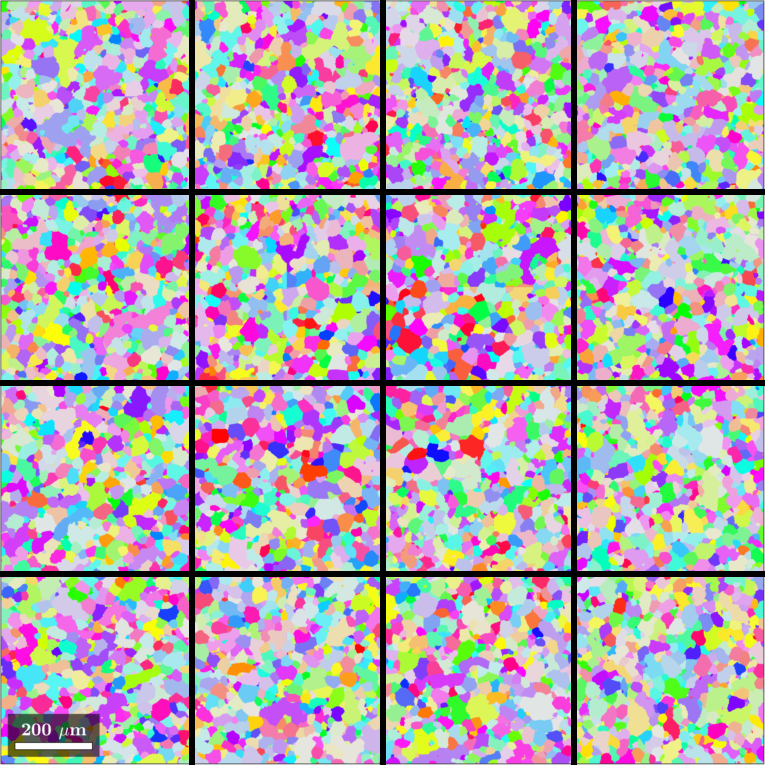}
    \includegraphics[width=0.49\linewidth]{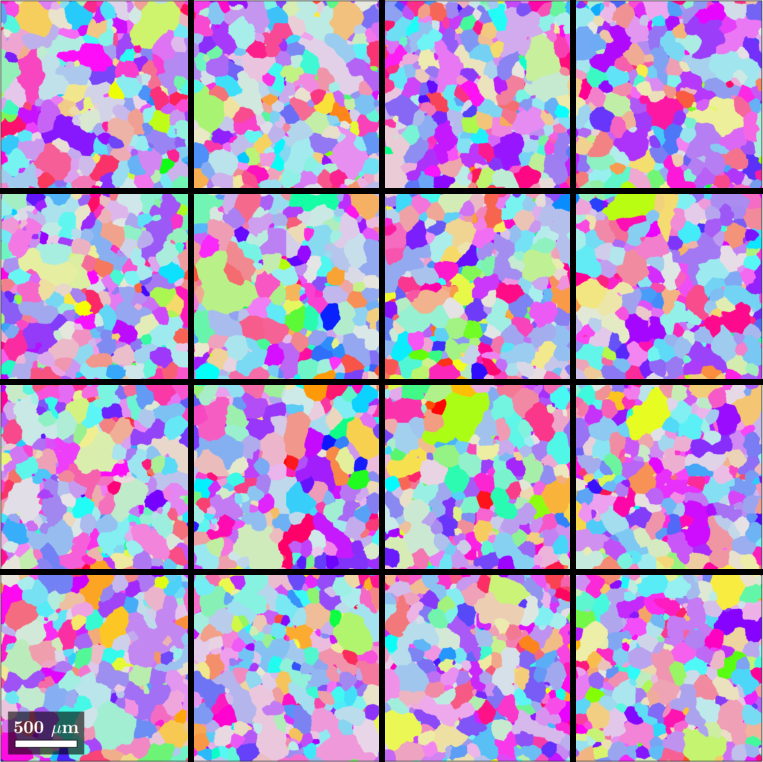}
    \caption{Visualization of local microstructural variability across the generated $2048\times2048$ EBSD maps (x4). Each large map (left: In718, right: In625) is divided into 16 non-overlapping $512\times512$ cells crops. These crops are used to evaluate spatial homogeneity by comparing their microstructural metrics distributions with those computed on the full generated map.}
    \label{fig:gridedmaps}
\end{figure}

For this analysis, as shown in Figure \ref{fig:gridedmaps}, we extracted $16$ non-overlapping $512\times512$ cells crops from each $2048\times2048$ extended map. For each crop, we computed the distribution of the key microstructural metrics described in Section \ref{dataprocessingmetrics}. We then visualized the minimum and maximum values observed across the 16 crops as vertical error bars for each bin, and overlaid the corresponding global distribution computed on the entire $2048\times2048$ map.

\begin{figure}[H]
    \centering
    \includegraphics[width=0.4\linewidth]{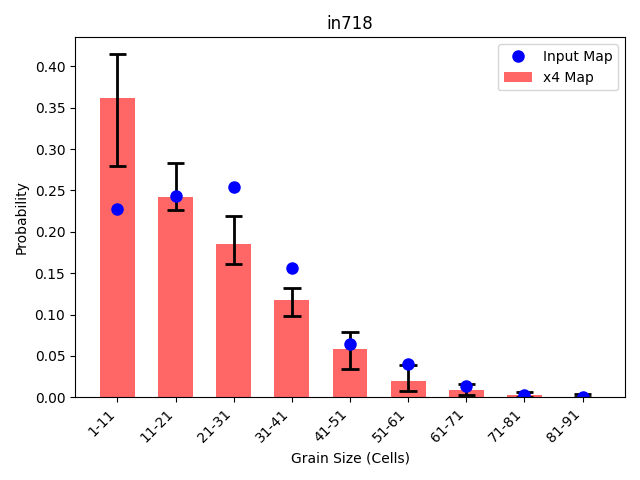}
    \includegraphics[width=0.4\linewidth]{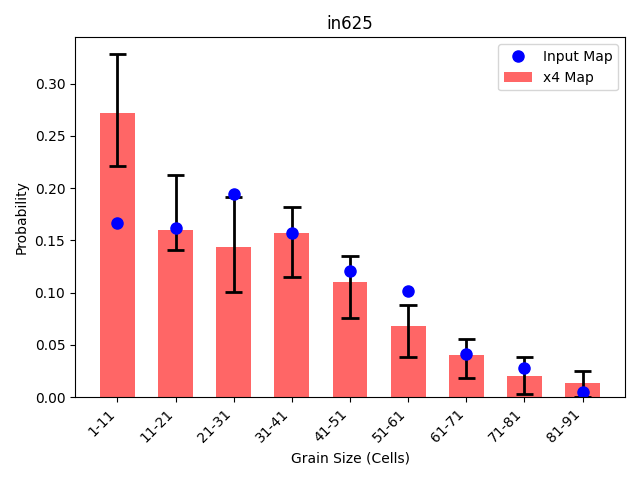}
    \caption{Number-weighted Grain Size Distributions comparison between the full $2048\times2048$ generated map (x4) (red bars) and its 16 non-overlapping $512\times512$ subregions (black error bars) with the error bars indicating the minimum and maximum values observed in the local crops.}
    \label{fig:grainsizehomo}
\end{figure}

\begin{figure}[H]
    \centering
    \includegraphics[width=0.4\linewidth]{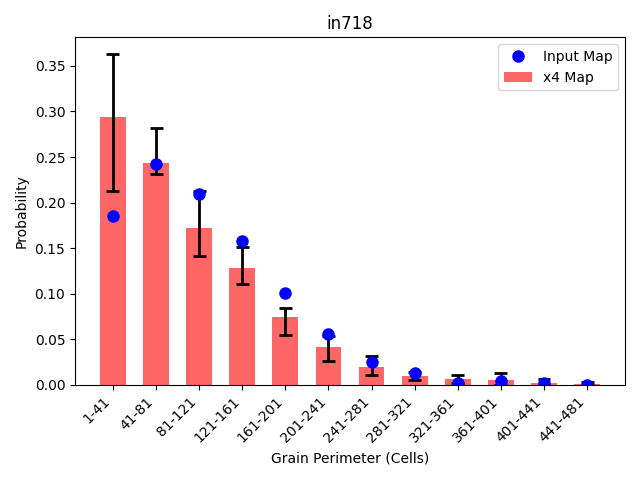}
    \includegraphics[width=0.4\linewidth]{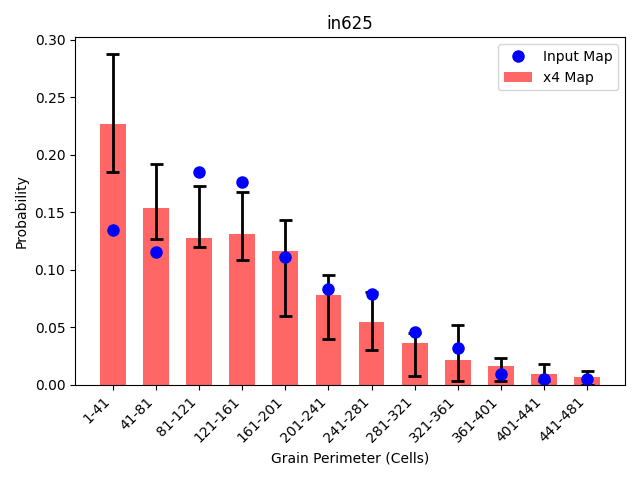}
    \caption{Grain Perimeter Distributions comparison between the full $2048\times2048$ generated map (x4) (red bars) and its 16 non-overlapping $512\times512$ subregions (black error bars) with the error bars indicating the minimum and maximum values observed in the local crops.}
    \label{fig:grainperimeterhomo}
\end{figure}

\begin{figure}[H]
    \centering
    \includegraphics[width=0.4\linewidth]{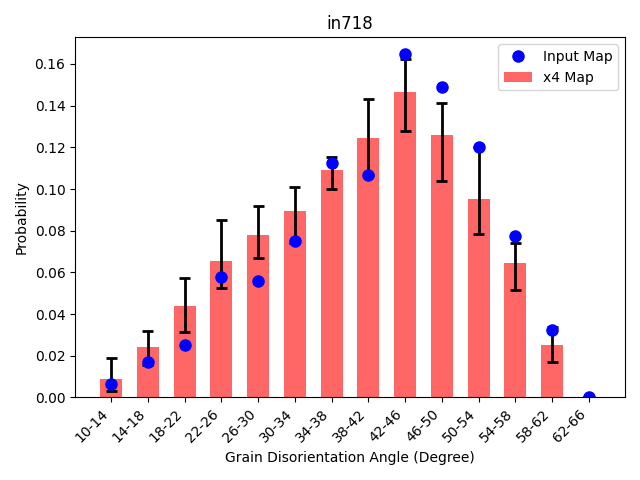}
    \includegraphics[width=0.4\linewidth]{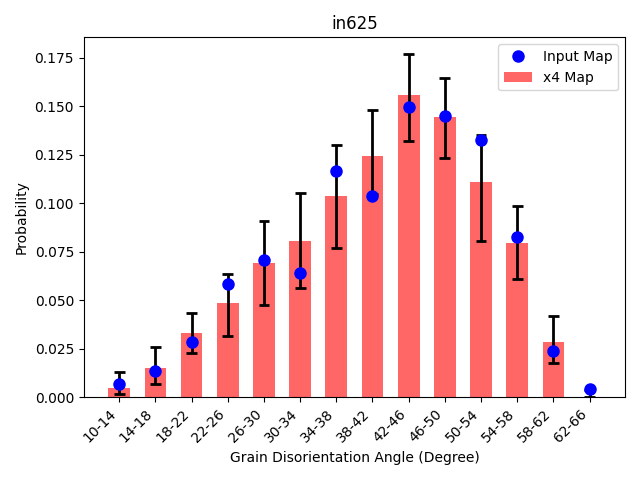}
    \caption{Grain Boundary Disorientation Distributions comparison between the full $2048\times2048$ generated map (x4) (red bars) and its 16 non-overlapping $512\times512$ subregions (black error bars) with the error bars indicating the minimum and maximum values observed in the local crops.}
    \label{fig:grainmisorientationhomo}
\end{figure}

\begin{figure}[H]
    \centering
    \includegraphics[width=0.4\linewidth]{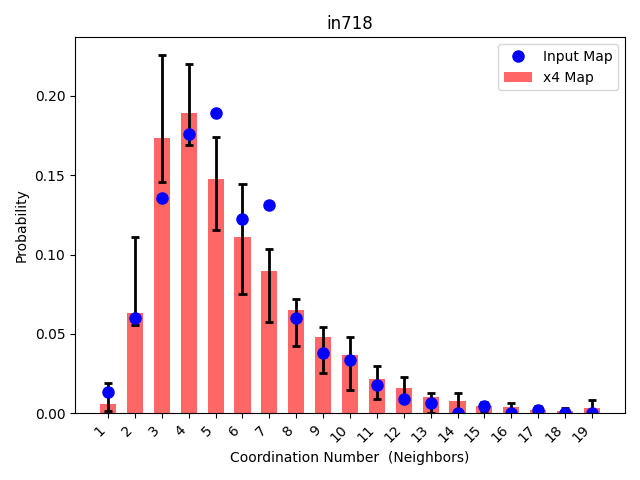}
    \includegraphics[width=0.4\linewidth]{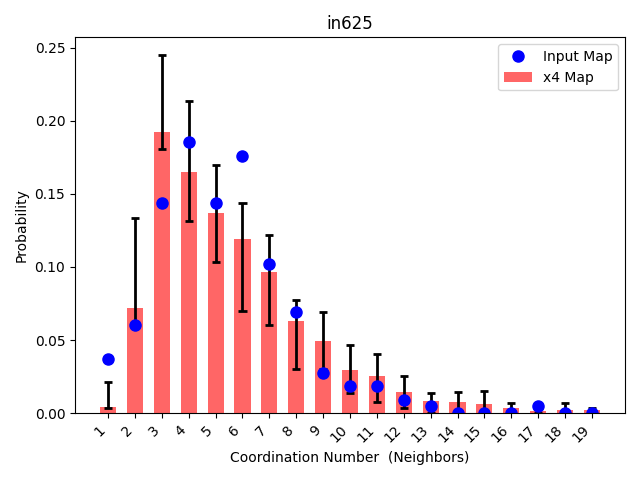}
    \caption{Grain Coordination Number Distributions comparison between the full $2048\times2048$ generated map (x4) (red bars) and its 16 non-overlapping $512\times512$ subregions (black error bars) with the error bars indicating the minimum and maximum values observed in the local crops.}
    \label{fig:grainneighborshomo}
\end{figure}

\begin{figure}[H]
    \centering
    \includegraphics[width=0.4\linewidth]{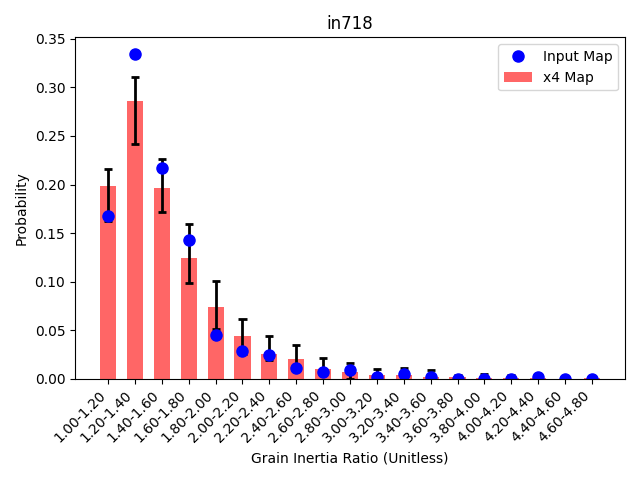}
    \includegraphics[width=0.4\linewidth]{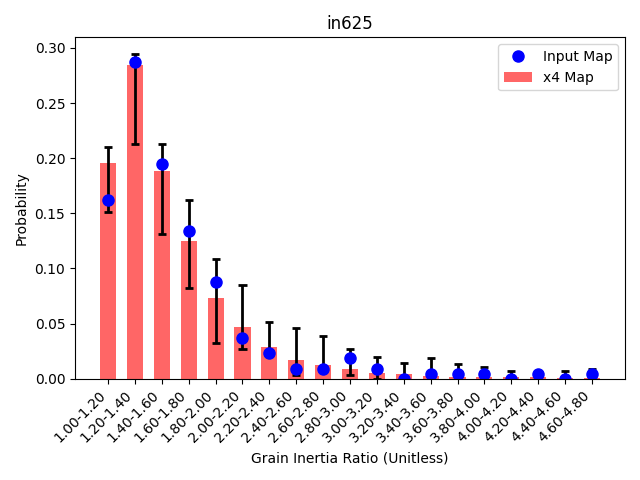}
    \caption{Grain Inertia Ratio Distributions comparison between the full $2048\times2048$ generated map (x4) (red bars) and its 16 non-overlapping $512\times512$ subregions (black error bars) with the error bars indicating the minimum and maximum values observed in the local crops.}
    \label{fig:inertiaratiohomo}
\end{figure}

Figures (\ref{fig:grainsizehomo}, \ref{fig:grainperimeterhomo}, \ref{fig:grainmisorientationhomo}, \ref{fig:grainneighborshomo}, \ref{fig:inertiaratiohomo}) reveal how much local variability exists between subregions and whether this variability remains within a plausible range. For both In718 and In625 examples, we observe that the majority of local distributions remain close to the global reference, with only moderate deviations in certain bins. This suggests that the generated maps exhibit good spatial homogeneity, and that the statistical characteristics are not localized or clustered, but instead well distributed across the entire synthetic map surface. Such an analysis provides additional confidence in the physical realism of the generated maps, as it demonstrates that the model does not introduce localized bias or artifacts during patch-wise generation.

\subsection{Generation from Metrics Only}
\label{resultsgeneration}

Beyond extending existing EBSD maps, the proposed method is also capable of generating entirely new microstructures from scratch, using only statistical conditioning metrics. This experiment highlights the model’s ability to synthesize spatially plausible EBSD maps that respect predefined statistical targets without any input map. 

\begin{figure}[H]
    \centering
    \includegraphics[width=0.6\linewidth]{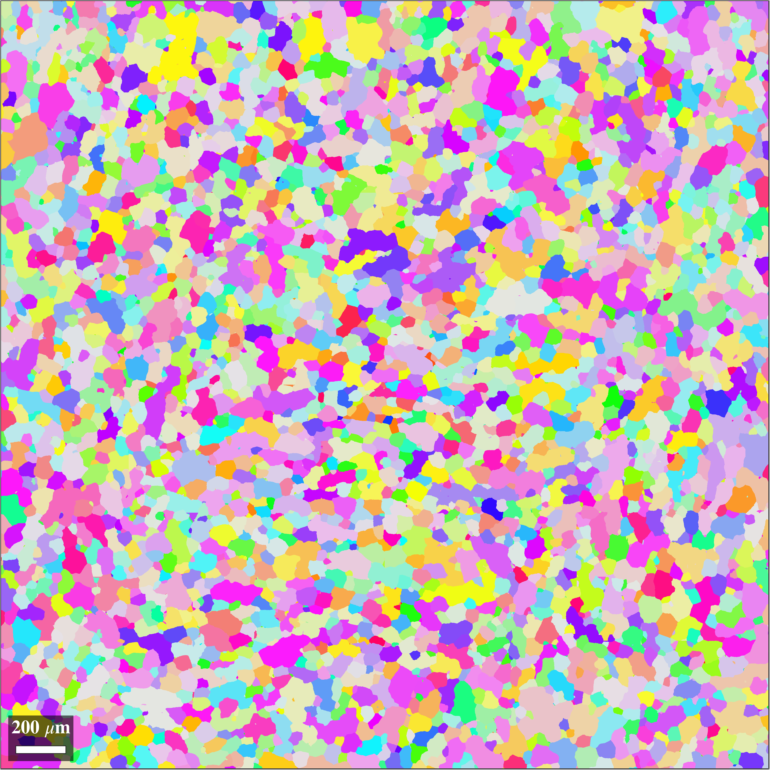}
    \caption{$3072\times3072$ Cells EBSD Map Generated without providing any input map. The model uses only the conditioning metrics to guide the generation.}
    \label{fig:generatedmap}
\end{figure}

\begin{figure}[H]
    \centering
    \includegraphics[width=0.48\linewidth]{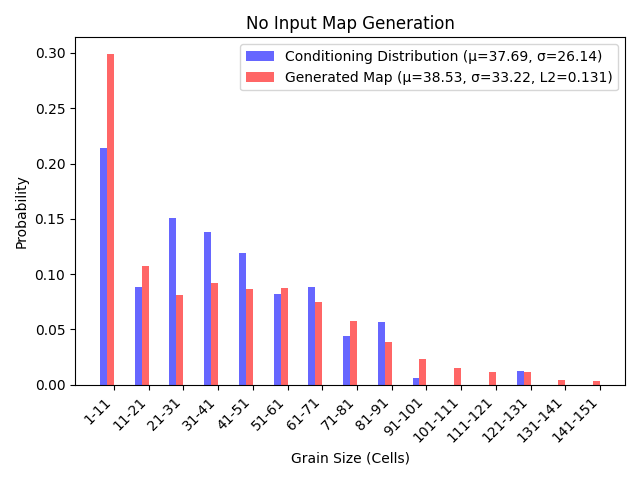}
    \includegraphics[width=0.48\linewidth]{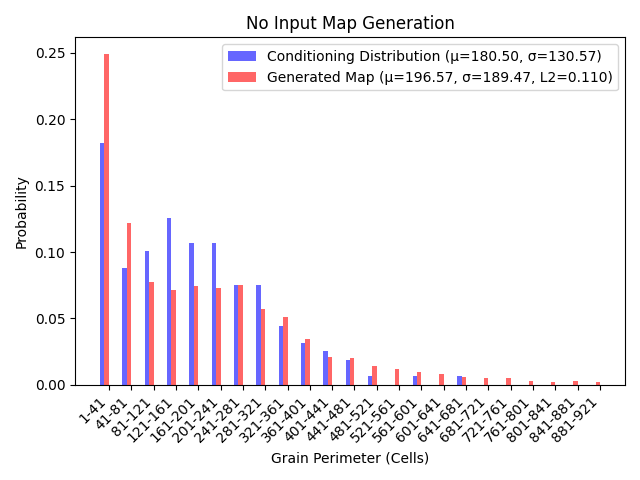}
    \caption{Comparison of number-weighted Grain Size and Perimeter Distributions between the conditioning vector (blue) and the generated EBSD map (red).}
    \label{fig:grainsizeperimeter}
\end{figure}

\begin{figure}[H]
    \centering
    \includegraphics[width=0.48\linewidth]{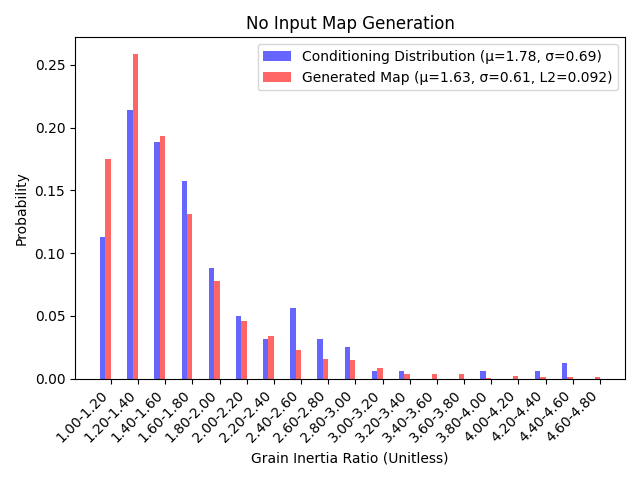}
    \includegraphics[width=0.48\linewidth]{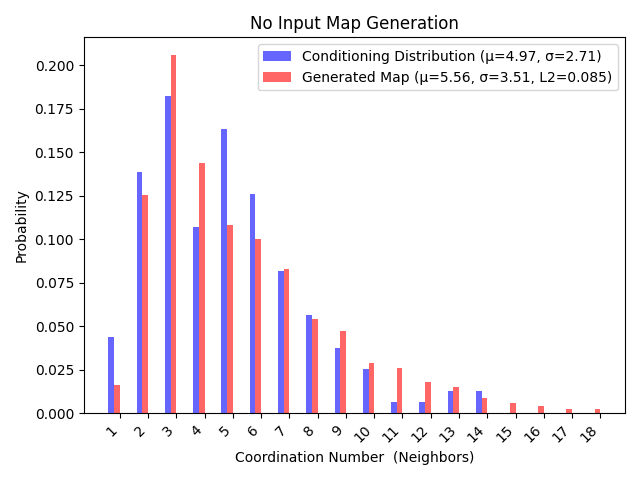}
    \caption{Comparison of Grain Inertia Ratio and Coordination Number Distributions between the conditioning vector (blue) and the generated EBSD map (red).}
    \label{fig:graininertiann}
\end{figure}

\begin{figure}[H]
    \centering
    \includegraphics[width=0.48\linewidth]{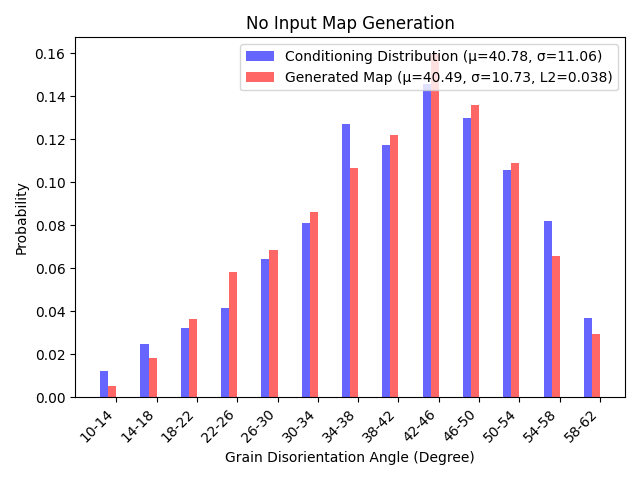}
    \caption{Comparison of Disorientation Angle Distributions between the conditioning vector (blue) and the generated EBSD map (red).}
    \label{fig:misorientationeuler1}
\end{figure}

Figure \ref{fig:generatedmap} shows a generated EBSD map of size $3072\times3072$ cells. The map exhibits realistic crystallographic and morphological patterns. To validate whether the generation respects the intended statistical targets, as shown in Figures (\ref{fig:grainsizeperimeter}, \ref{fig:graininertiann}, \ref{fig:misorientationeuler1}), we compute the metrics distributions from the generated map and compare it to the ones used as conditioning metrics. 

\subsection{Border Grains Analysis}
\label{resultsbordergrainsanalysis}

\begin{figure}[H]
    \centering
    \includegraphics[width=0.488\linewidth]{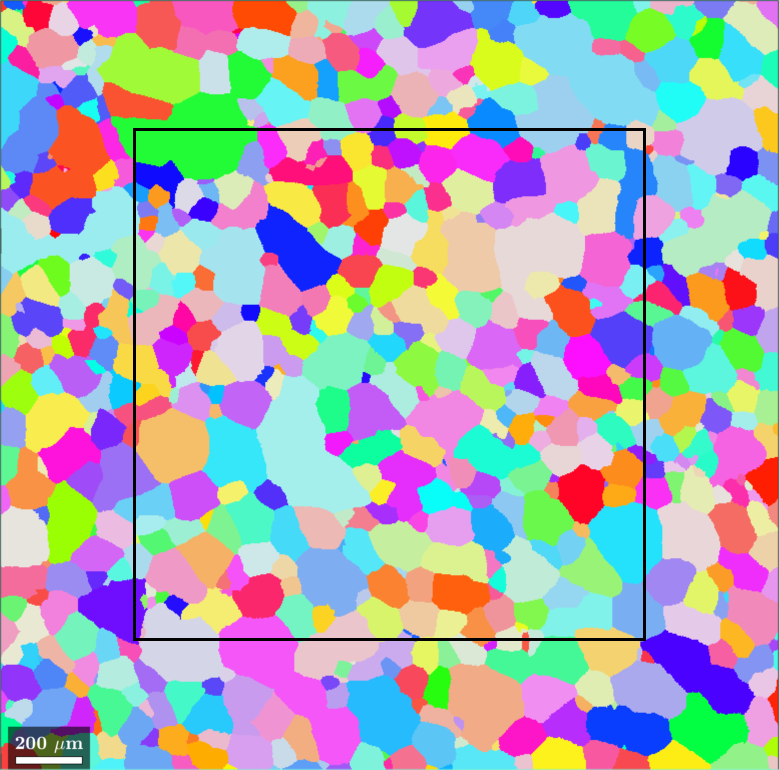}
    \includegraphics[width=0.488\linewidth]{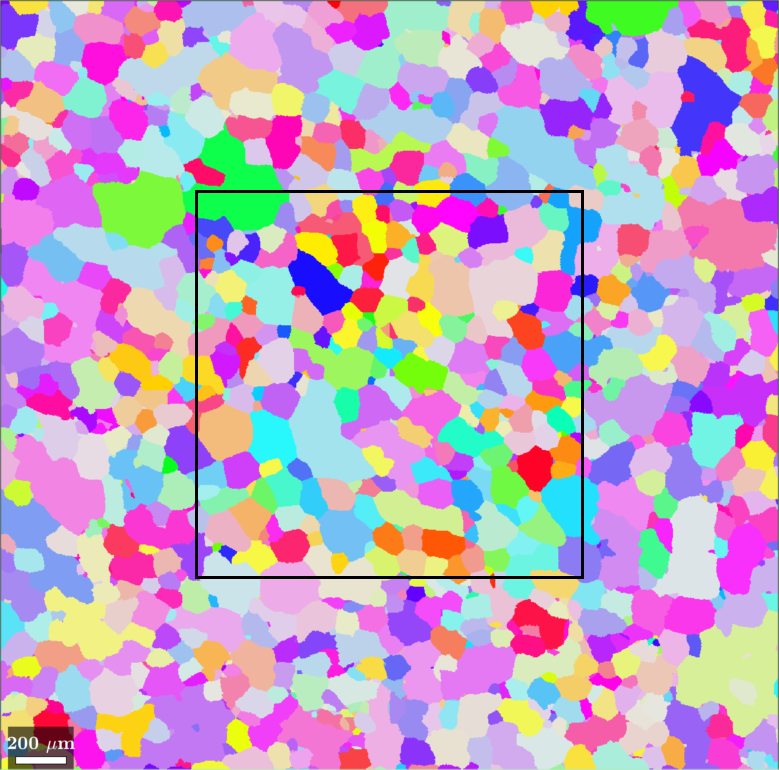}
    \caption{Left: Experimental $776\times776$ cells EBSD map with a $512\times512$ cells crop region shown in the center. Right: $1024\times1024$ cells extended EBSD map using the $512\times512$ cells crop from the experimental map as input. The central region corresponds to the crop; the surrounding areas are generated.}
    \label{fig:bordergrainsmaps}
\end{figure}

When extending an EBSD map, it is common for some grains at the borders of the input region to be only partially visible. A relevant question is whether the model can realistically complete these truncated grains during extension, producing physically plausible continuations in the synthesized regions. To explore this, we designed an experiment based on a real EBSD map of size $776\times776$ cells. From this map, we extracted a central crop of $512\times512$ cells, which served as the input for the extension with the model.  Figure \ref{fig:bordergrainsmaps} shows the original experimental map with the $512\times512$ crop outlined in the center. Next, the cropped input was extended to a size of $1024\times1024$ cells using the model. This generated map includes a central region which corresponds to the input map and the surrounding synthesized regions.

\begin{figure}[H]
    \centering
    \includegraphics[width=0.48\linewidth]{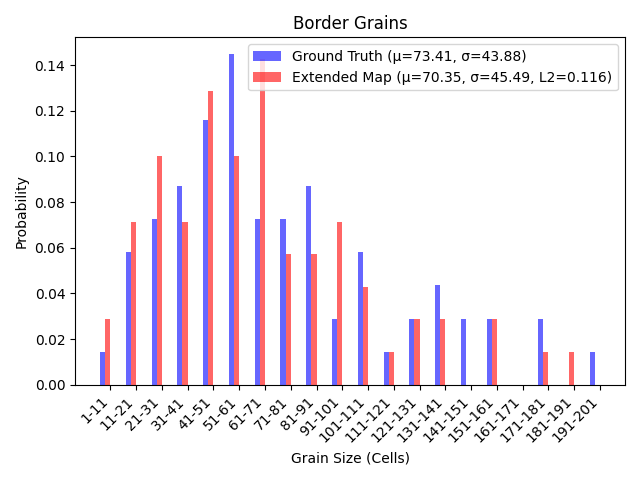}
    \includegraphics[width=0.48\linewidth]{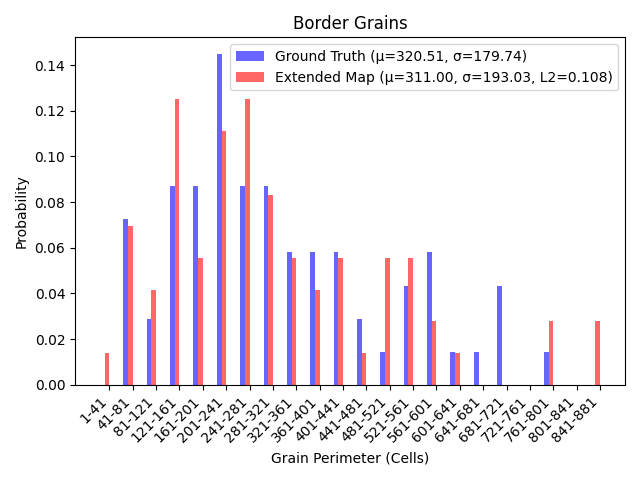}
    \caption{Comparison of Grain Size and Perimeter Distributions between ground-truth border grains (blue) and the grains extended in the synthesized regions (red).}
    \label{fig:bggrainsizeperimeter}
\end{figure}

\begin{figure}[H]
    \centering
    \includegraphics[width=0.48\linewidth]{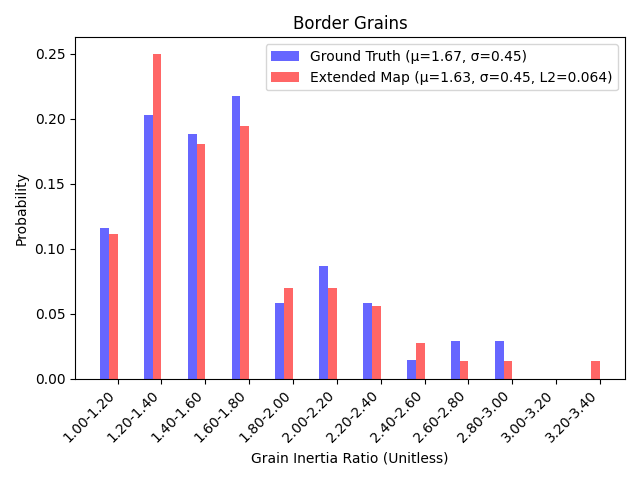}
    \includegraphics[width=0.48\linewidth]{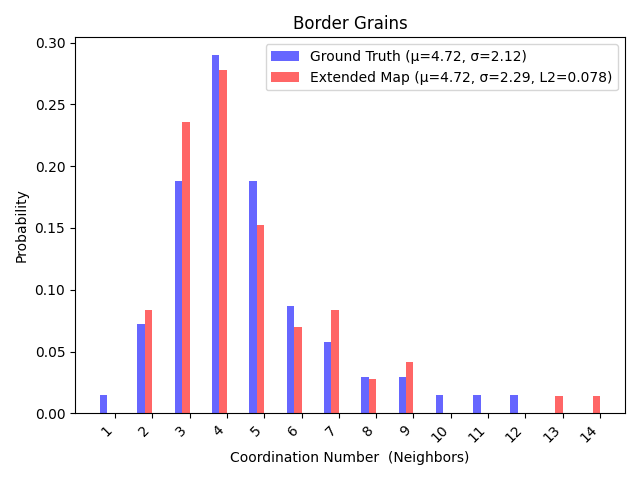}
    \caption{Comparison of Grain Inertia Ratio and Coordination Number Distributions between ground-truth border grains (blue) and the grains extended in the synthesized regions (red).}
    \label{fig:bggraininertiann}
\end{figure}

\begin{figure}[H]
    \centering
    \includegraphics[width=0.48\linewidth]{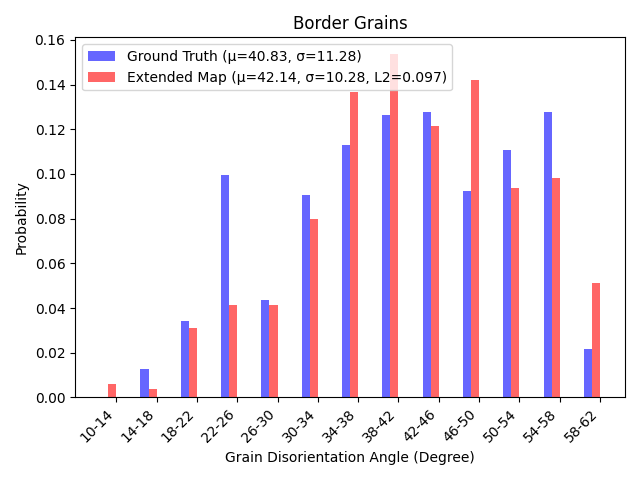}
    \caption{Comparison of Disorientation Angle Distributions between ground-truth border grains (blue) and the grains extended in the synthesized regions (red).}
    \label{fig:bgmisorientationeuler1}
\end{figure}

We then identified all grains in the original $776\times776$ map that are intersected by the border of this central crop. This yielded a set of approximately 70 border grains. To assess whether the model realistically completes the grains that were cut at the crop boundary, we compare two sets of grains: the set of border grains extracted from the full experimental map, these are grains that were partially present in the crop and fully visible in the experimental map. The set of border grains in the extended map, these are grains from the input map that lie on the border and extended by the model into the synthesized area. We compute and compare the distributions of key morphological and crystallographic metrics between these two groups. This allows us to evaluate whether the grains synthesized by the model at the crop borders resemble the natural grain continuations observed in experimental data. The distributions in Figures (\ref{fig:bggrainsizeperimeter}, \ref{fig:bggraininertiann}, \ref{fig:bgmisorientationeuler1}) show a strong overlap between the ground-truth and extended grains. These results indicate that the model does not extends arbitrarily, but instead meaningfully extends truncated grains in a physically consistent manner. This supports the robustness of the proposed patch-wise extension strategy, confirming its ability to handle grain continuation in a realistic and coherent way.

\section{Discussions}
\label{discussions}

The first key observation is that the model not only preserves morphological continuity across patch boundaries but also introduces new grain structures that differ from those seen in the input map, as shown in Figure \ref{fig:extendedmaps}. This suggests that the model does not simply replicate visible grains, but instead learns a statistical prior that allows it to synthesize novel, yet plausible, configurations. This ability is particularly important for applications where diversity and statistical representativeness are needed, such as metallurgical simulations in material science. In the case of generation from metrics alone, as shown in Figure \ref{fig:generatedmap}), the model successfully synthesizes large maps consistent with the specified targets. While minor fluctuations between the conditioning distributions and those measured from the output can be observed, these deviations remain within acceptable bounds. Thus, the results are highly promising and largely comparable to the error levels associated with Laguerre-Voronoï-type approaches which only allow grain size distributions to be matched using poorly realistic shapes, it nevertheless remains that the precision of the obtained results could still be improved. Two avenues are currently under investigation to assess their impact on reducing the observed $L_2$ errors across the different distributions: first, the influence of the data representation mode through distributions and bins of the raw data; and second, the fact that each morphological, topological, and crystallographic characteristic within the metric vector is assigned equal weight during the learning process.

The model demonstrates strong performance when generating microstructures whose statistical distributions fall within the range of those observed in the training dataset. As with any machine learning approach, the generalization capacity remains linked to the diversity and representativeness of the data used during training. Extending the model to handle exotic materials or distributions would require enriching the dataset with a broader variety of alloys, textures, and grain morphologies. 

Another important consideration lies in the condition of the EBSD maps used for training. In this study, the maps were pre-processed as described in Section \ref{dataprocessing}, using a mean orientation per grain, removing twins and applying a minimum grain size threshold to remove small artifacts, to simplify the learning process. Moreover, no second phase particle populations were considered. The model thus learns from filtered microstructures rather than raw EBSD acquisitions, which may contain features such as twins or dislocation density. But this approach still represents a breakthrough in the way microstructures can be generated for numerical simulations, offering a data-driven alternative to conventional synthetic generation methods used in simulation software such as DIGIMU\textsuperscript{\textregistered} \cite{Digimu,Digimu1,Bernacki2024}. The ability to generate large, statistically controlled EBSD maps opens up several use cases across materials science and engineering. One particularly promising application is the generation of representative microstructures from mean-field models predictions. In multiscale frameworks, where macroscopic simulations yield statistical descriptors of the microstructure (e.g., average grain size or orientation spread), our model can act as a downstream generator to reconstruct plausible microstructural fields consistent with these descriptors. Another use case lies in the creation of synthetic datasets for benchmarking in materials science research. The model can also assist in virtual material design by enabling the exploration of microstructure-property relationships. In all these cases, the ability to control specific statistical aspects of the generated maps, while ensuring their spatial realism, provides a powerful tool for both research and industrial simulation workflows.

A further point concerns the conditioning tensor $\mathbf{C}$, which currently expects eight distinct metrics as input. In practical applications, users might have access to only a subset of these metrics, often just the grain size distribution. In such cases, a reasonable strategy could involve retrieving the missing distributions from the dataset using similarity-based selection or clustering, ensuring that the provided grain size distribution remains dominant while the other metrics act as complementary priors. This flexibility would make the model more adaptable to real-world use cases, where partial microstructural descriptors are more common than fully characterized microstructural descriptors.

\section{Conclusion}
\label{conclusion}

InfinityEBSD introduces a new paradigm for microstructure generation, capable of producing large-scale EBSD maps of arbitrary size from either an initial input map of arbitrary size, or directly from user-specified statistical descriptors. By combining a latent diffusion model with metric-based conditioning and a patch-wise extension strategy, the method synthesizes spatially coherent and physically plausible microstructures for monophase polycrystalline materials. The generated maps are not only statistically consistent with the target metrics but also visually realistic and directly compatible with analysis tools.

Several directions remain open for future work as described previously. One of the first important step will involve removing current pre-processing constraints and training the model on raw EBSD maps that include twins and localized heterogeneities such as intragranular orientation gradients. This would enable a closer match to real-world microstructures as observed in experimental conditions. During training, microstructural metrics are computed on the full EBSD maps, including grains that intersect the map borders. While this approach ensures that all data is utilized, it may introduce slight distortions in the distributions due to partially visible grains. An interesting approach for future work would be to adapt the pre-processing pipeline to exclude border grains when computing metrics, both during training and evaluation. Another natural extension is to move beyond 2D generation: by learning from 2D slices and integrating dimensionality expansion techniques, the model could be adapted to synthesize plausible 3D microstructures, with applications in Representative Volume Element (RVE) generation.

\section*{Data Availability}

The dataset used in this study will be released separately in a dedicated publication.

\section*{CRediT authorship contribution statement}

\textbf{Sterley Labady:} Conceptualization, Methodology, Software, Validation, Formal Analysis, Investigation, Data Curation, Visualization, Writing - Original Draft. \textbf{Marc Bernacki:} Conceptualization, Validation, Supervision, Project Administration, Funding Acquisition, Writing - Review \& Editing. \textbf{Youssef Mesri:} Conceptualization, Validation, Resources, Supervision, Writing - Review \& Editing. \textbf{Daniel Pino Muñoz:} Conceptualization, Validation, Supervision, Writing - Review \& Editing. \textbf{Baptiste Flipon:} Software, Validation, Resources, Data Curation, Writing - Review \& Editing.

\section*{Declaration of competing interest} 
The authors declare that they have no known competing financial interests or personal relationships that could have appeared to influence the work reported in this paper. 

\section*{Acknowledgments} 
The authors thank ArcelorMittal, Aperam, Aubert \& Duval, CEA, Constellium, Framatome, Safran and the French National Research Agency (ANR) for their financial support through the DIGIMU\textsuperscript{\textregistered} consortium and RealIMotion ANR Industrial Chair (Grant No. ANR-22-CHIN-0003). 

The authors also wish to express their gratitude to Bowen Liu, PhD Researcher at CEMEF, for his constant availability and advices. 

\bibliography{biblio}

\end{document}